\documentclass[fleqn,usenatbib]{mnras}
\usepackage{newtxtext,newtxmath}
\usepackage[T1]{fontenc}
\usepackage{ae,aecompl}

\usepackage{graphicx,caption}
\usepackage{amsmath}

\usepackage{orcidlink}
\usepackage{graphicx}
\usepackage{xspace}

\newcommand\HI{H\textsc{i}\xspace}
\newcommand\HIMF{H\textsc{i}MF\xspace}
\newcommand\HIMFs{H\textsc{i}MFs\xspace}
\def\Sref#1{Sec.~\ref{#1}\xspace}
\def\Fref#1{Fig.~\ref{#1}\xspace}
\def\Eref#1{Eq.~\eqref{#1}\xspace}
\def\Tref#1{Table~\ref{#1}\xspace}

\usepackage{xcolor}

\title[\HI lensing]{Gravitational lensing of 21~cm \HI signal: detection prospects at $z\sim1$ with the uGMRT in galaxy cluster lenses}

\author[Bharti, Meena \& Bagla]{
Sauraj Bharti\orcidlink{0000-0001-9030-3926}$^{1}$\thanks{E-mail: \href{mailto:saurajbharti@gmail.com }{saurajbharti@gmail.com}},
Ashish Kumar Meena\orcidlink{0000-0002-7876-4321}$^{2,3}$,
and Jasjeet Singh Bagla\orcidlink{0000-0002-7749-4155}$^{1,4}$
\\
\\
$^{1}$Department of Physical Sciences, IISER Mohali, Knowledge City, Sector 81, SAS Nagar, Punjab 140306, India
\\
$^{2}$Department of Physics, Indian Institute of Science, Bengaluru 560012, India
\\
$^{3}$Physics Department, Ben-Gurion University of the Negev, P.O. Box 653, Be'er-Sheva 84105, Israel
\\
$^{4}$National Centre for Radio Astrophysics, Tata Institute of Fundamental Research, Ganeshkhind, Pune 411007, India
}

\date{Accepted XXX; Received YYY; in original form ZZZ}

\pubyear{2025}

\begin{document}
\label{firstpage}
\pagerange{\pageref{firstpage}--\pageref{lastpage}}
\maketitle

%%%%%%%%%%%%%%%%%%%%%%%%%%%%%%%%%%%%%%%%%%%%%%%%%%%%%%%%%%%%%%%%%%%%%%%%
\begin{abstract}
The atomic hydrogen \HI content of galaxies is intimately related to star formation and galaxy evolution through the baryon cycle, which involves processes such as accretion, feedback, outflows, and gas recycling. While probing the \HI gas over cosmic time has improved our understanding, direct \HI detection with the redshifted 21 cm line is essentially limited to $z\lesssim 0.42$.  Detections beyond this redshift are based on stacking to obtain average \HI mass of galaxy populations. Gravitational lensing by the cluster lenses enhances the \HI signal and can extend the redshift limit further. In this work, we describe simulations of \HI lensing in cluster lenses. We explore the feasibility of detecting strongly lensed \HI emission from background galaxies using known $50$ cluster lenses within the uGMRT sky coverage. We demonstrate that certain clusters offer a strong likelihood of \HI detection. We also investigate how strong lensing distorts the \HI spectral lines. The shape of the \HI signal in these lensing models provides useful information and can be used in optimising signal extraction in blind and targeted \HI surveys.  We find that blind detection of \HI signal from galaxies in the redshift range up to $1.58$ requires more than a few hundred hours of observations of individual clusters with the uGMRT. Detecting \HI emission in galaxies with strong optical lensing seems promising, with a $5\sigma$ detection potential in less than 50 hours for Abell 370 and 75 hours for Abell 1703 using the uGMRT.
\end{abstract}

%%%%%%%%%%%%%%%%%%%%%%%%%%%%%%%%%%%%%%%%%%%%%%%%%%%%%%%%%%%%%%%%%%%%%%%%

%%%%%%%%%%%%%%%%%%%%%%%%%%%%%%%%%%%%%%%%%%%%%%%%%%%%%%%%%%%%%%%%%%%%%%%%
\maketitle

%%%%%%%%%%%%%%%%%%%%%%%%%%%%%%%%%%%%%%%%%%%%%%%%%%%%%%%%%%%%%%%%%%%%%%%%
\section{Introduction}

Neutral hydrogen is the primary fuel for star formation, and it is responsible for galaxy evolution through a complex baryon cycle. 
The Baryon cycle collectively refers to the processes such as, conversion of atomic \HI gas into molecular gas $\rm{H_{2}}$, accretion, feedback mechanism, outflows, recycling of diffuse gas, which determine the formation and evolution of galaxies~\citep[e.g.,][]{Ford2014, Tumlinson2017, Oppenheimer2018}. 
Observations along with cosmological simulations (and semi-analytical models) have enhanced our understanding of the baryon cycle~\citep[e.g.,][]{Wright2024}, although limitations are still faced at sub-kilo-parsec scales \citep{Naab_2017}. 
To understand a complete picture of galaxy evolution, one needs to model the effects of the various processes~(and their interplay) involved in the baryon cycle, along with the distribution of ISM, the connection of the gas with internal and external properties of galaxies~\citep[e.g.,][]{Catinella2010, Saintonge2022}.

The detection of a large number of \HI galaxies with sufficient redshift coverage can help in studying gas distribution within the baryon cycle and its evolution over cosmic time. 
The \HI content of galaxies can be detected via absorption in the redshifted Lyman-$\alpha$ line of Hydrogen, as well as emission in the redshifted $21$cm \HI line originating from the hyperfine transition in the ground state of the hydrogen atom. 
The radiative transition probability of the $21$cm \HI line is very low. 
That said, since hydrogen is the most abundant element in the universe, the signal is detectable, although the direct detection is limited to the nearby Universe. 
With current radio facilities, direct \HI detections have been only possible up to $z=0.42$~\citep[e.g.,][]{Fernandez2016, Hongwei2024} and detecting the \HI signal at even higher redshifts requires stacking signals from multiple \HI galaxies, limiting us to study only the average properties~\citep[e.g.,][]{Rhee2013, Bera2019, Chowdhury2020, Chowdhury2021}.

Gravitational lensing offers a unique opportunity for direct detection of \HI signals from distant, faint sources. 
Gravitational lensing refers to the bending of light rays coming from a distant background source due to an intervening mass~(i.e., lens) distribution between the source and observer~\citep[e.g.,][]{1992grle.book.....S, 1996astro.ph..6001N}. 
Gravitational lensing can also produce multiple magnified images of the background, a regime known as strong lensing. 
The amount of magnification for a background source depends on its size and its distance from the caustics, which trace the high magnification regions in the source plane. 
For background galaxy sources detected at optical wavelengths, the maximum magnification factor is of the order of~$\sim10^2$. 
Since \HI distribution is more extended than its optical counterpart in galaxies, the magnification boost for \HI is typically expected to be lower than its optical counterpart. 
Even then, simulations have shown that to push the redshift limit to~$z\gtrsim1$ with the existing and upcoming facilities, the extra magnification boost provided by lensing seems to be sufficient~\citep[e.g.,][]{Deane_2015, 2025MNRAS.537.3134B} at least for the intrinsically bright \HI sources, i.e., galaxies with a high \HI mass. 
With the Giant Metrewave Radio Telescope~(GMRT), \citet{Blecher2019} looked at three galaxy-galaxy lenses and claimed a marginal detection of lensed \HI at~$z\sim 0.4$ in one of the lensed galaxies. Recently,~\citet{Chakraborty_2022} claimed \HI detection with upgraded-GMRT~(uGMRT) at~$z\sim1.3$ in a galaxy-galaxy lens system.
This claim has been contested by~\citet{2024MNRAS.535L..70D} on grounds of unphysical parameters inferred for the source galaxy. 

Galaxy clusters are the most massive strong gravitational lenses in the Universe~\citep[e.g.,][]{2011A&ARv..19...47K, 2024SSRv..220...19N} and are prime targets to search for lensed \HI sources due to their ability to produce highly magnified lensed images. 
Cluster lenses often have multiple background galaxies that are lensed with high magnification; hence, the chances of detection of strongly lensed sources are higher.  
Further, radio telescopes have large fields of view and hence the same observations can potentially be used for parallel studies that may justify long integration times. 
The upcoming Square Kilometre Array~(SKA) in phase-2 and its precursors will observe~$\sim3/4$ of the whole sky, which is expected to include~$\gtrsim 10^4$ galaxy clusters with masses~$\gtrsim 10^{14}\:{\rm M_\odot}$ \citep[e.g.,][]{Pillepich_2012}. 
Many of these clusters will lead to highly magnified lensed galaxies, pushing the direct \HI detection to~$z\gtrsim2$~\citep[e.g.,][]{Staveley-Smith:20151o, 2015MNRAS.450.2251Y}. 
Even with existing facilities, cluster lensing will allow us to push the \HI detection to~$z\sim 1$. 
For example, \citet{Blecher2024} looked at the Hubble Frontier Fields clusters to search for lensed galaxies with high \HI mass and high magnification factors and found that the `Dragon Arc' at redshift~0.725 in Abell~370~\citep{1987A&A...172L..14S} can lead to \HI detection in $\sim50$~hours with MeerKAT. 
Although the above encourages us to target and search for lensed \HI signals in galaxy cluster lenses, more studies are required to estimate the number of lensed \HI sources in large sky surveys with such lenses and their properties, such as redshift distribution, observed signal-to-noise ratio, i.e. SNR~(as well as detection likelihood), and lensed \HI line profile. 
Analysing the lensing effect on the \HI signal will allow us to devise better strategies to search for lensed \HI signals in blind surveys, as well as select targets with known multiply imaged systems in the optical, where high SNR is expected for the \HI component.

In our current work, we study the prospects of detecting lensed \HI signal in galaxy cluster lenses with uGMRT. 
We use a sample of fifty known galaxy cluster lenses in the uGMRT sky coverage with available strong lens models to study the properties of lensed \HI source and determine the fraction of systems that can be observed with 100 hours of integration time. 
In the absence of lensing, an \HI galaxy with non-zero inclination is expected to show a double-horn profile, detection of which allows us to determine source properties such as \HI mass, rotation velocities, and inclination. 
An asymmetry in the horns can be a sign of uneven \HI surface density distribution, perturbation in the galactic \HI disc, or tidal interaction~\citep[e.g.,][]{Bok_2018, Andersen_2009}. 
We demonstrate that the resolution limitation of radio telescope for distant galaxies can lead to asymmetries and distortions in the spectral line shape for galaxies just above the detection threshold.  
Strong lensing distorts the observed shape of the background source, leading to a large variety in the observed image configurations. 
Lensing-induced image distortions are expected to be large for a source located close to the caustics, which marks the regions of high magnification in the source plane. 
This implies that a highly magnified \HI source will have relatively more complicated image formation, which will, in turn, also distort the observed \HI line profile~\citep{Deane_2015}. 
Hence, we also study the effect of strong gravitational lensing on the observed \HI line profile to understand its impact on the detection of \HI signal and its SNR.

The current work is organised as follows. 
\Sref{sec:gl_basics} briefly reviews the relevant basics of gravitational lensing. 
\Sref{sec:gc_sample} discusses the galaxy cluster sample used in the current work. 
The simulation method to generate mock (lensed) source populations is described in \Sref{sec:mock_sim}. 
The method to estimate (lensed) source SNR is outlined in \Sref{sec:snr}. 
Results are presented in \Sref{sec:results}. 
We conclude and summarise our work in \Sref{sec:conclusions}. 
Throughout this work, we assume a flat $\Lambda$CDM cosmology described by, $(\Omega_m, \Omega_{\Lambda}, h) = (0.3, 0.7, 0.7)$.

%%%%%%%%%%%%%%%%%%%%%%%%%%%%%%%%%%%%%%%%%%%%%%%%%%%%%%%%%%%%%%%%%%%%%%%%
\section{Basics of gravitational lensing}
\label{sec:gl_basics}

In this section, we briefly review the relevant basics of gravitational lensing~\citep[e.g.,][]{1992grle.book.....S, 1996astro.ph..6001N}.  Gravitational lensing of a background source by a foreground mass distribution (i.e., lens), assuming the thin lens approximation, is described by the gravitational lens equation, 
\begin{equation}
    \pmb{\eta} = \pmb{\chi} - \pmb{\zeta}(\pmb{\chi}),
    \label{eq:lens_eq}
\end{equation}
where~$\pmb{\eta}$ and~$\pmb{\chi}$ represent the angular position of the unlensed source in the source plane and its images in the image (or lens) plane, respectively. $\pmb{\zeta}(\pmb{\chi})$ is the deflection angle and related to the 2D projected lens potential~($\psi$) as,~$\pmb{\zeta}(\pmb{\chi}) = \nabla\psi(\pmb{\chi})$. The properties of the observed images (such as their observed shape and magnification) can be described by the corresponding Jacobian matrix,
\begin{equation}
    \mathbb{A}(\pmb{\chi}) \equiv \frac{\partial\pmb{\eta}}{\partial\pmb{\chi}} = \delta_{ij} - \psi_{ij}
\end{equation}
where subscripts denote the partial derivatives with respect to~$\pmb{\chi}$ components, i.e., $\psi_i = \partial\psi/\partial\chi_i$. We can also write down the above equation in a matrix form,
\begin{equation}
    \mathbb{A}(\pmb{\chi}) = 
    \begin{pmatrix}
        1 - \kappa - \gamma_1 & -\gamma_2 \\
        -\gamma_2 & 1 - \kappa + \gamma_1
    \end{pmatrix},
\end{equation}
where~$\kappa$ and $\gamma = \gamma_1 + \iota \gamma_2$ represent the well-known convergence and shear at the image position, respectively. The convergence controls the isotropic distortion of the image, whereas the shear stretches/compresses the image in a particular direction. The magnification of an observed image, assuming a point source, is given as
\begin{equation}
    \mu \equiv \frac{1}{\det\mathbb{A}} = \frac{1}{(1-\kappa)^2 - \gamma^2}.
\end{equation}
As we can see, magnification goes to infinity at certain points in the image plane depending on the~$(\kappa, \gamma)$ values. Such points form smooth closed curves in the image plane known as critical curves, and the corresponding closed curves (not necessarily smooth) in the source plane are known as caustics. The total magnification of a point source is given by,
\begin{equation}
    \mu_{\rm p} (\pmb{\eta}) = \sum_{i=1}^n \frac{1}{\left|(1-\kappa_i)^2 - \gamma_i^2\right|},
    \label{eq:mu_point}
\end{equation}
where~$n$ is the total number of images. In our current work, since we investigate lensing of extended \HI sources, the resulting magnification is given by the surface brightness weighted mean over the source area ~\citep[e.g., see chapter~7 in][]{1992grle.book.....S},
\begin{equation}
    \mu_{\rm T}(\pmb{\eta}) = \frac{\int \mu_{\rm p}(\pmb{\eta}') \: I(\pmb{\eta}'-\pmb{\eta}) \: d^2\pmb{\eta}'}{\int I(\pmb{\eta}') \: d^2\pmb{\eta}'} \simeq \frac{\sum \mu_p(\pmb{\eta}) \: I(\pmb{\eta}'-\pmb{\eta})}{\sum I(\pmb{\eta}')},
    \label{eq:mu_ext}
\end{equation}
where~$I(\pmb{\eta})$ is the source brightness profile. As we pixelate our sources, we use the above approximation for the corresponding magnification estimations.

%%%%%%%%%%%%%%%%%%%%%%%%%%%%%%%%%%%%%%%%%%%%%%%%%%%%%%%%%%%%%%%%%%%%%%%%
\begin{table}
    \centering
    \caption{Parameters of various \HIMF used in this work. We combine the \HIMF parameters from \citet{Bera_2022} and \citet{Chowdhury_2024} to construct a \emph{combined} \HI mass function. See \Sref{sec:mock_sim} for more details.}
    \begin{tabular}{lccc}
    \hline
    \HIMF     & $\alpha$ & $\log ({\rm M^*_{\HI}} )$ & $\phi^{*}$ \\
    \hline
    \citet{Martin_2010}    &  -1.33   &  9.96   & $4.81\times10^{-3}$ \\
    \citet{Bera_2022}      &  -1.29   &  9.60   & $12.44\times10^{-3}$ \\
    \citet{Chowdhury_2024} &  -1.25   & 10.14   & $12.61\times10^{-3}$ \\
    \hline
    \end{tabular}
    \label{tab:HIMF}    
\end{table}  
%%%%%%%%%%%%%%%%%%%%%%%%%%%%%%%%%%%%%%%%%%%%%%%%%%%%%%%%%%%%%%%%%%%%%%%%

%%%%%%%%%%%%%%%%%%%%%%%%%%%%%%%%%%%%%%%%%%%%%%%%%%%%%%%%%%%%%%%%%%%%%%%%
\begin{table*}
    \centering
    \caption{Sample of galaxy cluster lenses used in this work. Columns~(2), (3), and~(4) represent the cluster name and its position in (RA, Dec), respectively. The cluster redshift is given in column~(5). The lens models, their resolutions, and the underlying survey names for each cluster used in the current work are shown in columns ~(6), (7), and (8), respectively. For \textsc{Lenstool} models, we use the ones constructed by the Sharon team~\citep[e.g.,][]{2014ApJ...797...48J}.}
    \begin{tabular}{rlllllll}
    \hline
    \#  &  Cluster name      & RA         & Dec         & $z_l$    & Lens model               & Resolution    & Catalogue    \\
    (1) &     (2)            & (3)        & (4)         & (5)      & (6)                      & (7)           & (8)          \\
    \hline    
    1.  & Abell 370          & 39.9704167 &  -1.5768056 & 0.375    & \textsc{Lenstool}        & 0.050$''$     & HFF          \\
    2.  & Abell 2744         & 3.5883333  & -30.3972500 & 0.308    & \textsc{Lenstool}        & 0.050$''$     & HFF          \\
    3.  & Abell S1063        & 342.185000 & -44.5301389 & 0.348    & \textsc{Lenstool}        & 0.050$''$     & HFF          \\
    4.  & MACSJ0416.1-2403   & 64.0370833 & -24.0746389 & 0.396    & \textsc{Lenstool}        & 0.050$''$     & HFF          \\
    5.  & MACSJ0717.5+3745   & 109.391666 &  37.7469444 & 0.543    & \textsc{Lenstool}        & 0.050$''$     & HFF          \\
    6.  & MACSJ1149.5+2223   & 177.401250 &  22.3994722 & 0.545    & \textsc{Lenstool}        & 0.050$''$     & HFF          \\
    7.  & Abell 1763         & 203.828750 &  40.9992222 & 0.228    & \textsc{glafic}          & 0.100$''$     & RELICS       \\
    8.  & Abell 2163         & 243.951250 &  -6.1268611 & 0.203    & \textsc{glafic}          & 0.100$''$     & RELICS       \\
    9.  & Abell 2537         & 347.092500 &  -2.1923333 & 0.297    & \textsc{glafic}          & 0.100$''$     & RELICS       \\
    10. & Abell 2813         & 10.8545833 & -20.6207778 & 0.292    & \textsc{Lenstool}        & 0.100$''$     & RELICS       \\
    11. & Abell 3192         & 59.7212500 & -29.9291111 & 0.425    & \textsc{glafic}          & 0.100$''$     & RELICS       \\
    12. & Abell 697          & 130.745416 &  36.3641944 & 0.282    & \textsc{glafic}          & 0.100$''$     & RELICS       \\
    13. & CLJ0152.7-1357     & 28.1787500 & -13.9586111 & 0.833    & \textsc{glafic}          & 0.100$''$     & RELICS       \\
    14. & MACS J0025.4-1222  & 6.3762500  & -12.3800278 & 0.586    & \textsc{glafic}          & 0.100$''$     & RELICS       \\
    15. & MACS J0035.4-2015  & 8.8625000  & -20.2611944 & 0.352    & \textsc{glafic}          & 0.100$''$     & RELICS       \\
    16. & MACS J0159.8-0849  & 29.9558333 &  -8.8333333 & 0.405    & \textsc{glafic}          & 0.100$''$     & RELICS       \\
    17. & MACS J0257.1-2325  & 44.2925000 & -23.4366111 & 0.505    & \textsc{glafic}          & 0.100$''$     & RELICS       \\
    18. & MACS J0308.9+2645  & 47.2320833 &  26.7602222 & 0.356    & \textsc{glafic}          & 0.100$''$     & RELICS       \\
    19. & MACS J0417.5-1154  & 64.3904167 & -11.9062778 & 0.443    & \textsc{glafic}          & 0.100$''$     & RELICS       \\
    20. & MS1008.1-1224      & 152.640000 & -12.6619444 & 0.306    & \textsc{Lenstool}        & 0.100$''$     & RELICS       \\
    21. & PLCK G171.9-40.7   & 48.2370833 &  8.3720000  & 0.270    & \textsc{glafic}          & 0.100$''$     & RELICS       \\
    22. & PLCK G287.0+32.9   & 177.711666 & -28.0811667 & 0.390    & \textsc{glafic}          & 0.100$''$     & RELICS       \\
    23. & RXC J0600.1-2007   & 90.0408333 & -20.1358056 & 0.460    & \textsc{glafic}          & 0.100$''$     & RELICS       \\
    24. & RXC J0911.1+1746   & 137.797500 &  17.7759722 & 0.505    & \textsc{glafic}          & 0.100$''$     & RELICS       \\
    25. & RXC J0949.8+1707   & 147.462083 &  17.1209167 & 0.383    & \textsc{glafic}          & 0.100$''$     & RELICS       \\
    26. & RXC J2211.7-0350   & 332.941250 &  -3.8290833 & 0.397    & \textsc{glafic}          & 0.100$''$     & RELICS       \\
    27. & RXS J060313.4+4212N& 90.8008333 &  42.2568611 & 0.228    & \textsc{glafic}          & 0.100$''$     & RELICS       \\
    28. & RXS J060313.4+4212S& 90.8566667 &  42.1648889 & 0.228    & \textsc{glafic}          & 0.100$''$     & RELICS       \\
    29. & WHL J24.3324-8.477 & 24.3541667 &  -8.4569444 & 0.566    & \textsc{glafic}          & 0.100$''$     & RELICS       \\
    30. & Abell 1423         & 179.322340 &  33.6109625 & 0.213    & \textsc{Zitrin-NFW}      & 0.065$''$    & CLASH         \\
    31. & Abell 209          & 22.9689339 & -13.6112129 & 0.206    & \textsc{Zitrin-NFW}      & 0.065$''$    & CLASH         \\
    32. & Abell 2261         & 260.613235 &  32.1324784 & 0.224    & \textsc{Zitrin-NFW}      & 0.065$''$    & CLASH         \\
    33. & Abell 383          & 42.014090  &  -3.5292641 & 0.187    & \textsc{Zitrin-NFW}      & 0.065$''$    & CLASH         \\
    34. & Abell 611          & 120.23674  &  36.056565  & 0.288    & \textsc{Zitrin-NFW}      & 0.065$''$    & CLASH         \\
    35. & CLJ1226+3332       & 186.742667 &  33.5468250 & 0.890    & \textsc{Zitrin-NFW}      & 0.065$''$    & CLASH         \\
    36. & MACS J0329-02      & 52.4232238 &  -2.1962170 & 0.450    & \textsc{Zitrin-NFW}      & 0.065$''$    & CLASH         \\
    37. & MACS J0429-02      & 67.4000461 &  -2.8851911 & 0.399    & \textsc{Zitrin-NFW}      & 0.065$''$    & CLASH         \\
    38. & MACS J0647+70      & 101.958458 &  70.2471389 & 0.591    & \textsc{Zitrin-NFW}      & 0.065$''$    & CLASH         \\
    39. & MACS J0744+39      & 116.219987 &  39.4573883 & 0.686    & \textsc{Zitrin-NFW}      & 0.065$''$    & CLASH         \\
    40. & MACS J1115+01      & 168.966267 &   1.4986290 & 0.352    & \textsc{Zitrin-NFW}      & 0.065$''$    & CLASH         \\
    41. & MACS J1206-08      & 181.55065  &  -8.8009395 & 0.440    & \textsc{Zitrin-NFW}      & 0.065$''$    & CLASH         \\
    42. & MACS J1311-03      & 197.757519 &  -3.1777071 & 0.494    & \textsc{Zitrin-NFW}      & 0.065$''$    & CLASH         \\
    43. & MACS J1423+24      & 215.949486 &  24.0784633 & 0.545    & \textsc{Zitrin-NFW}      & 0.065$''$    & CLASH         \\
    44. & MACS J1720+35      & 260.069797 &  35.6073103 & 0.391    & \textsc{Zitrin-NFW}      & 0.065$''$    & CLASH         \\
    45. & MACS J1931-26      & 292.956795 & -26.5757730 & 0.352    & \textsc{Zitrin-NFW}      & 0.065$''$    & CLASH         \\
    46. & MACS J2129-07      & 322.358583 &  -7.6913333 & 0.570    & \textsc{Zitrin-NFW}      & 0.065$''$    & CLASH         \\
    47. & MS 2137.3-2353     & 325.063170 & -23.6611312 & 0.313    & \textsc{Zitrin-NFW}      & 0.065$''$    & CLASH         \\
    48. & RXJ 1347-1145      & 206.877543 & -11.7526358 & 0.451    & \textsc{Zitrin-NFW}      & 0.065$''$    & CLASH         \\
    49. & RXJ 2129+0005      & 322.416469 &   0.0892109 & 0.234    & \textsc{Zitrin-NFW}      & 0.065$''$    & CLASH         \\
    50. & RXJ 2248-4431      & 342.183243 & -44.5308625 & 0.348    & \textsc{Zitrin-NFW}      & 0.065$''$    & CLASH         \\
    \hline
    \end{tabular}
    \label{tab:clusters}    
\end{table*}  
%%%%%%%%%%%%%%%%%%%%%%%%%%%%%%%%%%%%%%%%%%%%%%%%%%%%%%%%%%%%%%%%%%%%%%%%

%%%%%%%%%%%%%%%%%%%%%%%%%%%%%%%%%%%%%%%%%%%%%%%%%%%%%%%%%%%%%%%%%%%%%%%%
\section{Galaxy cluster sample}
\label{sec:gc_sample}
In our current work, we use a total of fifty cluster lenses, all lying within the uGMRT sky coverage range. Out of these fifty galaxy clusters, six are from the \emph{Hubble Frontier Fields}~\citep[HFF\footnote{\url{https://archive.stsci.edu/prepds/frontier/}};][]{Lotz_2017} survey program, twenty-three are from the \emph{Reionization Lensing Cluster Survey}~\citep[RELICS\footnote{\url{https://archive.stsci.edu/hlsp/relics}};][]{2019ApJ...884...85C} program, and the remaining twenty-one are from the \emph{Cluster Lensing And Supernova survey with Hubble}~\citep[CLASH\footnote{\url{https://www.stsci.edu/~postman/CLASH/index.html}};][]{2012ApJS..199...25P} program. For various lensing purposes, we use the corresponding parametric lens models constructed using the \textsc{Lenstool}~\citep{2007NJPh....9..447J, 2011ascl.soft02004K}, \textsc{glafic}~\citep{2010ascl.soft10012O}, and \textsc{Zitrin-NFW}~\citep{2015ApJ...801...44Z} methods. The relevant details for each cluster are provided in \Tref{tab:clusters}.

Based on the quality of lensing data (and on the judgment of different teams), the available lensing data products for different cluster lenses vary in resolution from~$0.05''$ to~$0.1''$. However, the uGMRT sky resolution for sources at~$0.4 \lesssim z \lesssim 1.58$ is much lower ($\sim 3''-5''$) than the resolution of lensing data products. Hence, the non-homogeneity in the resolution of lensing data products is not expected to have an impact on our results. We note that this set of clusters is a biased sample and in terms of virial mass, the most massive clusters are over-represented in comparison to clusters with masses below $10^{14.8}$~M$_\odot$.

%%%%%%%%%%%%%%%%%%%%%%%%%%%%%%%%%%%%%%%%%%%%%%%%%%%%%%%%%%%%%%%%%%%%%%%%
\section{Mock Source Population}
\label{sec:mock_sim}
We simulate sources in volume between the given cluster lens redshift ($z_l$) and the maximum source redshift,~$z_s=1.58$. This upper limit is set by the frequency coverage of uGMRT Band-4. To draw redshifts of these sources, we use the differential comoving volume as a weight. The positions of these sources are chosen randomly within a cylinder with a comoving diameter of $10$~Mpc. A $10$~Mpc region is set to populate the sources behind each cluster as we find this to contain the region from which galaxies can be strongly lensed by clusters of galaxies in the catalog. All sources beyond three times the Einstein radius are then dropped from further consideration and we process only those within the strong-lensing region. To assign \HI masses to each of these sources, we assume that the \HI mass function~(\HIMF) follows the well-known Schechter function~\citep{Schechter}, which is given as,
\begin{equation}
    \phi(M_{\rm \HI}) = \ln(10) \: \phi^* \left( \frac{M_{\rm \HI}}{M_{\rm \HI}^*} \right)^{\alpha+1}
                        \exp \left(- \frac{M_{\rm \HI}}{M_{\rm \HI}^*} \right),
\end{equation}
where $\alpha$ is the lower mass end slope, $M_{\rm \HI}^*$ is knee mass in unit of $\rm M_{\odot}$ and $\phi^*$ (per $\rm Mpc^{3}$) is the normalization. We take three different sets of values for $(\alpha, M_{\rm \HI}^*, \phi^*)$, given in \Tref{tab:HIMF}, and then construct two \HI mass functions, which we call as, 
\begin{itemize}
    \item ALFALFA \HIMF: for which Schechter parameters $(\alpha, M_{\rm \HI}^*, \phi^*)$ are taken from~\citet{Martin_2010} where the authors directly measured the \HIMF using data from the ALFALFA survey \citep{Giovanelli_2005}. Although parameter values were derived using galaxies at~$z<0.06$, we use the corresponding \HIMF uniformly throughout the redshift range.
    
    \item Combined \HIMF: for which we use Schechter parameter $(\alpha, M_{\rm \HI}^*, \phi^*)$ values from~\cite{Bera_2022} for sources at~$z<1.0$, and parameter values from~\cite{Chowdhury_2024} for sources at $z\geq1.0$. These studies used the $M_{\rm HI}$-$M_{\rm B}$ scaling relation to estimate the \HIMF. These scaling relations were derived from \HI stacking of star-forming galaxies (in bins of B-band magnitude) at $z\approx 0.20-0.42$ in the Extended Growth Strip~\citep[EGS;][]{Zhao_2009} and at $z=0.74-1.45$ in the Deep2 field~\citep{Coil_2004}, respectively, which was then combined with the B‐band luminosity function to estimate the \HIMF. The upper cutoff in \HI masses is taken $10^{10.5}$ $\rm M_{\odot}$ for the sources sampled from both \HIMFs.
\end{itemize}

Once we have the redshift and \HI mass of a background galaxy source, the next step is to model the projected \HI distribution in the source plane. To compute the \HI disk size of the background galaxy, we use \HI size-mass~($D_{\rm \HI} - M_{\rm \HI}$) relation from~\citet{Wang__2025}. Recent studies have shown that this relation does not evolve with redshift and holds true for a wide range of source galaxy morphologies~\citep[e.g.,][]{Lelli2016, Gault2021, Rajohnson2022}. In general, the \HI distribution of a galaxy is much more extended than the optical light. To model the extended \HI surface density, we adopt a recently derived analytical approximation of the \HI profile by \citet{Wang__2025}. This approximate profile is derived based on the radial distribution of \HI down to surface density of~$0.01 \rm M_{\odot}\,pc^{-2}$, i.e., $R_{001}$ in the images obtained by the FEASTS\footnote{\url{https://github.com/FEASTS/LVgal/wiki}} program~\citep{2025ApJ...984...15Y}  and given as,
\begin{equation}
    y = \log \frac{(1+(x/r_{c})^2)^{-\beta}}{100(1+(1/r_{c})^2)^{-\beta}},
    \label{eq:HI_prof}
\end{equation}
where $y= \log(\Sigma_{\rm HI}/\rm M_{\odot}\,pc^{-2})$, $x = r/R_{001}$, $r_c = 0.94^{+0.09}_{-0.07}$ and $\beta = 8.43^{+1.15}_{-0.87}$. The scaling relation between $R_{001}$ and $M_{\rm \HI}$ is given as,
\begin{equation}
    \log{R_{001}} = 0.49(\pm0.02) \log{M_{\rm \HI}}-3.11(\pm0.21).
    \label{eq:r001}
\end{equation}
We use the best-fit values for various parameters and ensure consistency for total mass within~$R_{001}$\footnote{Parameters in the fitting formula for surface density are varied within the allowed range to ensure that we recover the \HI mass in Eq.~\eqref{eq:r001} used to compute $R_{001}$.}. The above setup gives us a circular source, whereas actual sources will have random inclination~($i$) and orientation~($\phi$) on the sky. For each source, we draw random~$(i, \phi)$ values such that $\cos{(i)}$ is uniformly distributed in $[0,1]$ and $\phi$ is uniformly distributed in $[0, 2\pi]$. For each source, we also simulate the stellar light profile to compare the \HI and optical magnifications. We use a truncated S\'{e}rsic profile~\citep{1963BAAA....6...41S} to model the stellar light of the source, which is given as,
\begin{equation}
\Sigma(r) = \Sigma_{0} \left[ \exp\left(-\left(\frac{r}{r_s}\right)^{1/n}\right) - \exp\left(-\left(\frac{r_t}{r_s}\right)^{1/n}\right) \right],
\label{eq:optical_prof}
\end{equation}
where $\Sigma_{0}$ denotes the central density, $r_s$ is the scale radius, and $r_t$ represents the truncation radius. To determine various parameters of the optical surface density profile, we follow these steps:
\begin{itemize}
    \item As most of the \HI sources are expected to be spiral galaxies, we set the S\'{e}rsic index~($n$) equal to one.
    
    \item To determine the stellar mass of our source, we use the relation between \HI mass and stellar mass given in Equation~(8) of~\citet{2018ApJ...864...40P}.
    
    \item Following~\citet{2025PASA...42...46L}, the truncation radius~($r_t$) is set equal to the stellar disk radius~($\rm R25$), which represents the isophotal radius where the i-band surface brightness is equal to $25\, \rm mag\,arcsec^{-2}$, and expected to contain $70\%$ of the total \HI mass. \citet{2025PASA...42...46L} also showed that the mean densities of \HI within stellar disc ($\Sigma_{\rm HI,R25}$) and within \HI disc ($\Sigma_{\rm HI,R_{\rm \HI}}$) are equal to~$3.04 \,\rm M_{\odot}\,pc^{-2}$ and~$2.42\,\rm M_{\odot}\,pc^{-2}$, respectively, where~$R_{\rm \HI}$ denotes the radius at which \HI surface density is falls to~$1 \rm M_{\odot}\,pc^{-2}$. To estimate the values of both $R25$ and $R_{\rm \HI}$, we solve \Eref{eq:HI_prof} with the above constraints on the mean densities assuming a relative scatter of 25\% and 20\% for $\Sigma_{\rm HI,R25}$ and $\Sigma_{\rm HI,R_{\rm HI}}$, respectively.
    
    \item For typical spiral galaxies, the scale radius is~$2-3$~kpc, which is $\sim15\%$ of the truncation radius, which is also what we adopt in our simulations, i.e., $r_s = 0.15 \times r_t$.
    
    \item The central density~($\Sigma_0$) is calculated such that the total mass inside the~$r_t$ is equal to the $90\%$ stellar mass of the galaxy, as $\rm R25$ is expected to contain $90\%$ of the total stellar light.
\end{itemize}
Here, it is important to note that we use the optical source truncation radius~($r_t$) such that the corresponding i-band surface brightness is equal to~$25\, \rm mag\,arcsec^{-2}$ based on observations in the local Universe~($z\lesssim0.08$). As we consider sources at~$z\simeq1$, due to the decrease in the observed flux, the observed source size will also decrease. However, for simplicity, we simulate optical sources up to~$r_t$ at all redshifts to estimate the optical magnification. By doing so, we are expected to underestimate the optical magnifications. However, since we only use optical magnification for comparison with \HI magnification, it does not affect any of our \HI results.

%%%%%%%%%%%%%%%%%%%%%%%%%%%%%%%%%%%%%%%%%%%%%%%%%%%%%%%%%%%%%%%%%%%%%%%%
\begin{figure}
    \centering
    \includegraphics[width=1.0\linewidth]{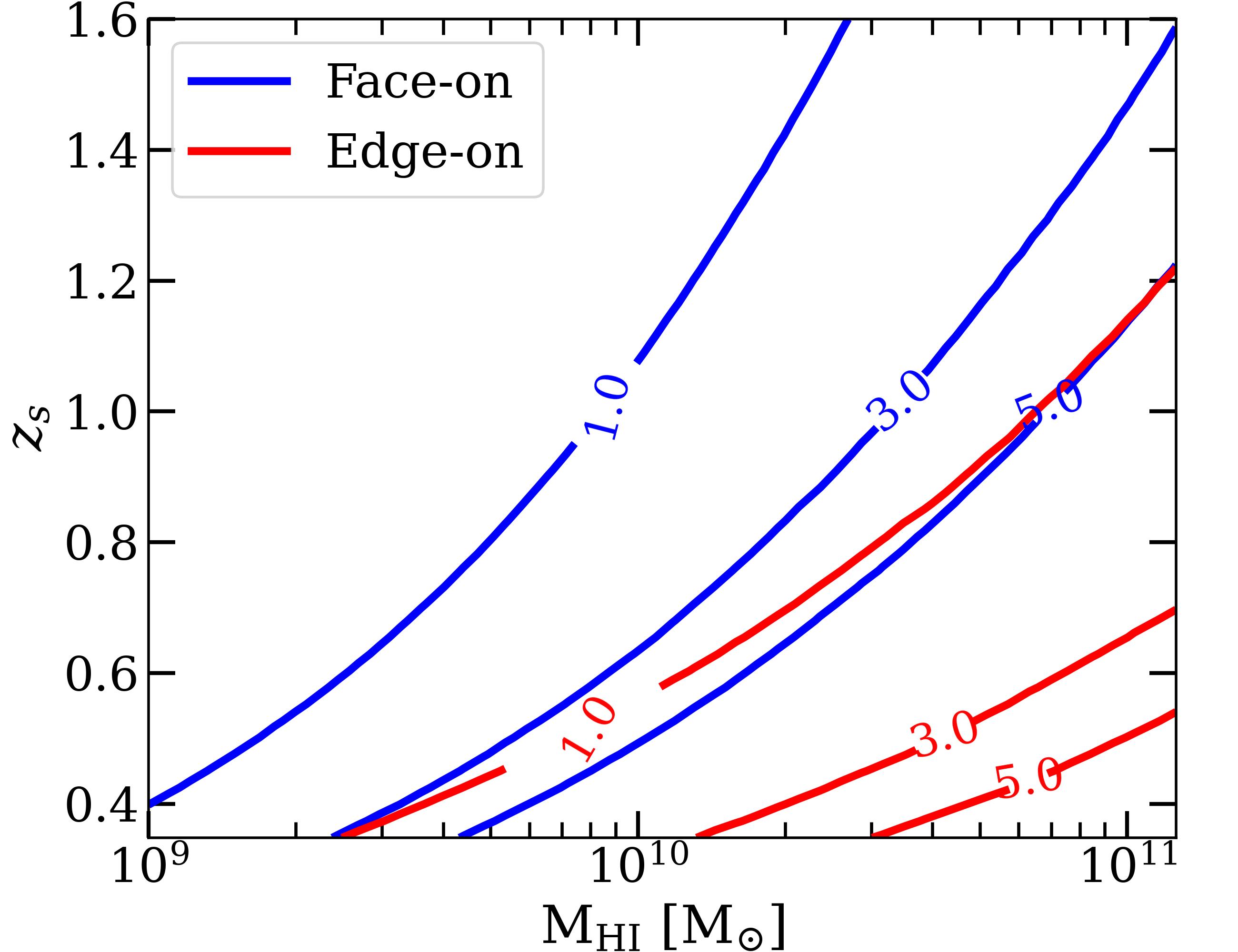}
    \caption{SNR dependence on the source inclination (without lensing). The blue and red contours are constant SNR curves for face-on and edge-on galaxies, respectively. We can see that the face-on \HI galaxies require less magnification compared to the edge-on galaxies, as the narrow linewidth has a relatively high peak flux density for the face-on galaxies.}
    \label{fig:Incl_effect}
\end{figure}
%%%%%%%%%%%%%%%%%%%%%%%%%%%%%%%%%%%%%%%%%%%%%%%%%%%%%%%%%%%%%%%%%%%%%%%%

%%%%%%%%%%%%%%%%%%%%%%%%%%%%%%%%%%%%%%%%%%%%%%%%%%%%%%%%%%%%%%%%%%%%%%%%
\begin{figure*}
    \centering
    \includegraphics[width=0.8\linewidth]{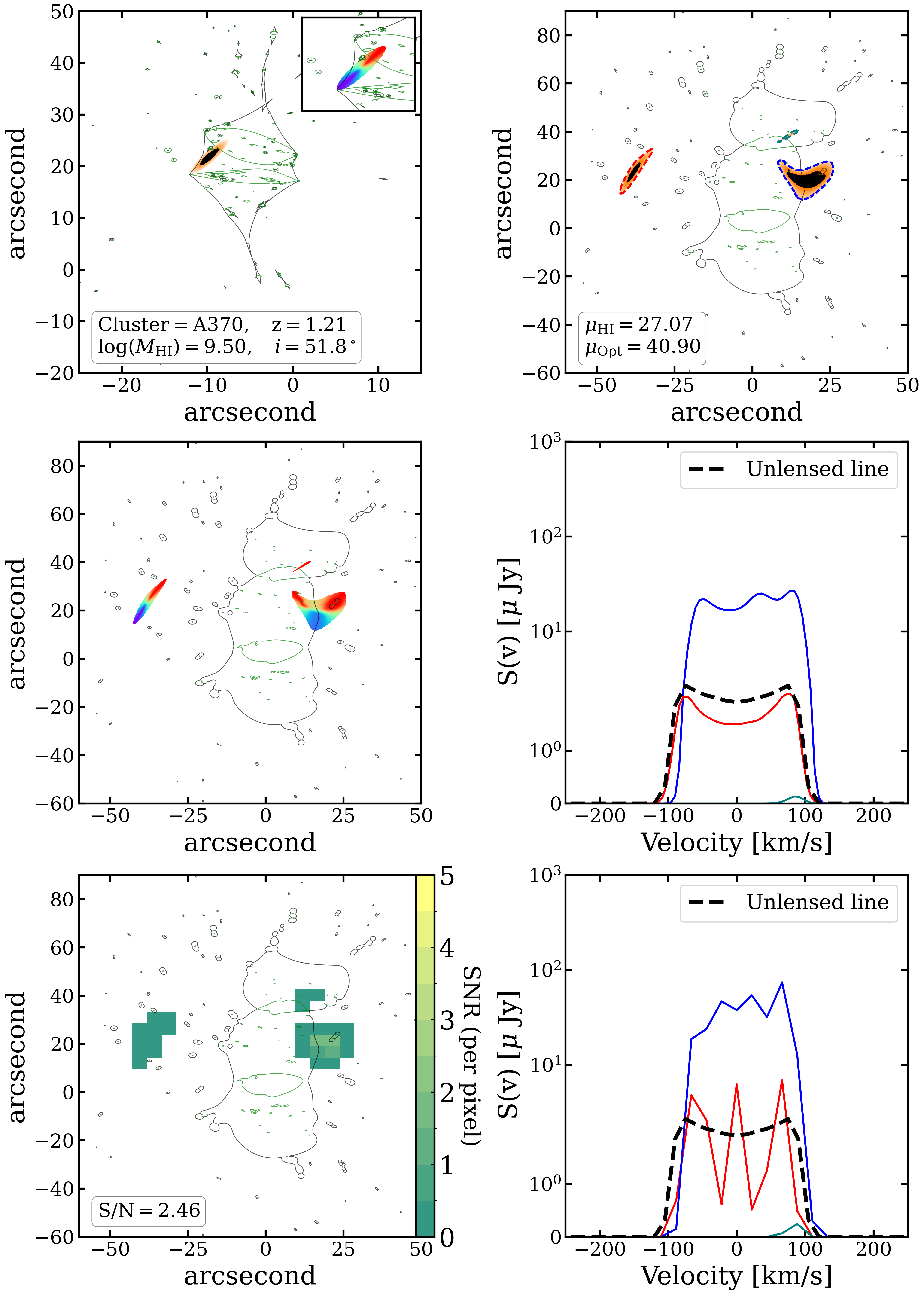}
    \caption{An example of lensed \HI source in Abell~370 and its SNR estimation at uGMRT resolution. \emph{Top-left panel} shows the source plane with black and green curves showing the tangential and radial caustics, respectively. The stellar density profile is shown by a black disk shape, and the orange region around it shows the \HI surface density extent. The inset plot shows the \HI velocity profile with red/blue marking red/blue-shifted regions. \emph{Top-right panel} shows the image plane with black and green curves representing the tangential and radial critical curves, respectively. \emph{Middle-left panel} shows the \HI velocity distribution in the lensed \HI images. \emph{Middle-right panel} shows the \HI line profile assuming the sky resolution is the same as the lens map resolution. The black dashed curve represents the \HI line profile of the unlensed source. The green, blue, and red curves show the lensed \HI line profile corresponding to the three lensed images enclosed in the same color contour in the top-right panel. \emph{Bottom-left panel} shows the image plane map of integrated SNR as seen by the uGMRT. Note that the uGMRT pixel size is much larger~($\sim3''-5''$) compared to the lensing map pixel size~($0.05''$). \emph{Bottom-right panel} shows the lensed \HI line profiles as seen by uGMRT on the coarser grid shown in the bottom-left panel.}
    \label{fig:snrmaps}
\end{figure*}
%%%%%%%%%%%%%%%%%%%%%%%%%%%%%%%%%%%%%%%%%%%%%%%%%%%%%%%%%%%%%%%%%%%%%%%%

%%%%%%%%%%%%%%%%%%%%%%%%%%%%%%%%%%%%%%%%%%%%%%%%%%%%%%%%%%%%%%%%%%%%%%%%
\section{SNR calculation}
\label{sec:snr}

%%%%%%%%%%%%%%%%%%%%%%%%%%%%%%%%%%%%%%%%%%%%%%%%%%%%%%%%%%%%%%%%%%%%%%%%
\subsection{Unlensed source SNR}
\label{ssec:unlensed_snr}
To calculate the SNR for an unlensed source, we start by estimating the effective number of baselines~($N_B$) of the uGMRT telescope that contribute at the required resolution. For this purpose, we simulate the UV coverage for the given declination (see \citealt{Bharti_2022} for more details). Here, we want to calculate the optimal SNR for a given source. Hence, we only choose those baselines for which the source is just resolved, since using longer baselines will over-resolve the source, potentially leading to a loss in the overall signal. For a given source size, $\theta_{\rm size}$, the effective baselines are those for which $B\leq B_{\rm crit}$ with~$B_{\rm crit}$ representing the baseline at which source size matches the angular resolution of the array, i.e.,
\begin{equation}
    \theta_{\rm size} = 1.22\frac{\lambda_{\rm obs}}{B_{\rm crit}},
    \label{eq:src_size}
\end{equation}
where $\lambda_{\rm obs}$ is the observed wavelength. With the effective number of baselines~($N_B$) calculated, the expected thermal noise~\citep{Meyer} in the receiver system~(in Jy/synthesized beam) is given as,
\begin{equation}
    \sigma_{\rm rms}=\frac{(T_{\rm sys}/G)}{\sqrt{2N_{B}\,\Delta t\,\Delta \nu}},
    \label{eq:noise}
\end{equation}
where $T_{\rm sys}$ is the system temperature, $\Delta t$ is the integration time, $G$ represents the antenna gain, and $\Delta \nu$ is the frequency width over which the flux of the source is spanned. A larger frequency width optimises detection sensitivity by capturing the entire spectral line over a bandwidth. Following parameters for the Band-4 ($550$ MHz to $850$ MHz) of uGMRT, we use $(G, T_{\rm sys})=(0.35~{\rm K/Jy}, 100~{\rm K})$. The expected \HI signal flux density ($S_{v}$) of assuming an optically thin source at redshift $z$, with \HI mass $M_{\rm HI}$ is given as,
\begin{equation}
    S_{v} = \frac{\beta(\theta)}{(2.356\times\,10^5\,W_{20})}\frac{M_{\rm HI}\,(1+z)}{D^2_L}
    \label{eq:signal}
\end{equation}
where $D_L$ is the luminosity distance to the source, $\beta(\theta)$ is the primary beam pattern of the antenna and $W_{20}$ is the linewidth of the source. We use the Baryonic Tully-Fisher Relation~\citep[BTFR;][]{McGaugh_2000} to estimate the circular velocity of the source, and with the given inclination, derive the linewidth $W_{20}$ at $20\%$ of the peak flux. \Eref{eq:signal} along with \Eref{eq:noise} are used to estimate the final SNR for a given unlensed source.

%%%%%%%%%%%%%%%%%%%%%%%%%%%%%%%%%%%%%%%%%%%%%%%%%%%%%%%%%%%%%%%%%%%%%%%%

\begin{figure*}
    \centering
    \includegraphics[width=0.9\linewidth]{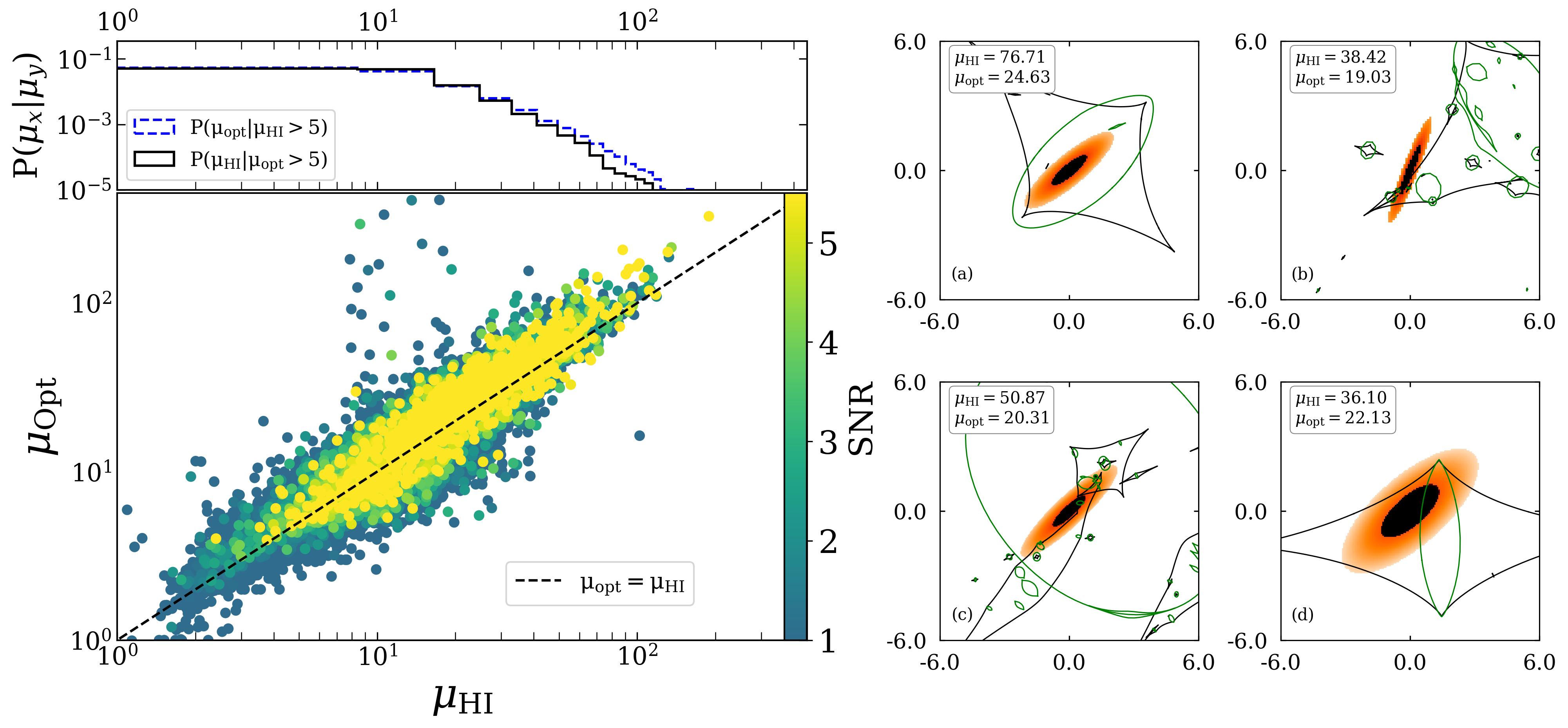}
    \caption{ Optical and \HI magnification comparison of sources from $50 \times 200$ simulations~(i.e., 200 realisations for each of the 50 clusters). The log–log scatter plot compares the optical and \HI\ magnifications for simulated sources. Each point is color-coded by its SNR, with higher-SNR points plotted on top to show the expected number and distribution of high-SNR sources. The corresponding conditional probability density distribution,~$P(\mu_x|\mu_y)$, of optical and \HI magnification is shown in the histogram above. On the right side, we present four cut-outs to highlight the source plane for cases where the \HI magnification is larger than the corresponding optical magnification (see \Sref{ssec:opt_vs_HI} for details).}    
    \label{fig:mu_distribution}
\end{figure*}
%%%%%%%%%%%%%%%%%%%%%%%%%%%%%%%%%%%%%%%%%%%%%%%%%%%%%%%%%%%%%%%%%%%%%%%%

In observations, for an \HI source with non-zero inclination, the spectral line profile is symmetric and double-horn, highlighting the symmetric nature of \HI density distribution, and absence of perturbations in the velocity field. To model the velocity field of an unlensed \HI source, we assume a $\tanh$ rotation curve with the corresponding scale radius, $r_s$, set equal to~$15\%$ of $R_{001}$, given as,
\begin{equation}
    V(r) = V_c\tanh{\left( \frac{r}{r_s} \right)}\,\frac{x}{r}\,\sin{i},
    \label{eq:Vfield}
\end{equation}
where $i$ is the inclination of the source and $V_c$ is the maximum rotation velocity of the source, estimated using the BTFR. $r$ is the radial distance from the source centre and $x$ is its projection on the x-axis, i.e.,~$x=r\cos\theta$. Given the \HI density profile and the rotation curve, the observed flux density in a velocity channel,~$(v, v+dv)$, can be given as~\citep[e.g.,][]{Gordon_1971,Schulman_1994,2105.04570v1},
\begin{equation}
    S(v) = {\rm Const.} \int^{R_{001}}_0 2\:\pi\:r\:\Sigma_{\rm HI}(r)\,dr\int^{\pi}_0 \mathcal{G}(v-V(r),\sigma_{v})\,d\theta,
    \label{eq:HIline}
   \end{equation}
where~$\mathcal{G}(v-V(r),\sigma_{v})$ is a Gaussian function given as,
\begin{equation}
    \mathcal{G}(v-V(r),\sigma_{v})\, = \frac{1}{\sqrt{2\pi}  \sigma_v} \exp\left( -\frac{(v - V(r))^2}{2 \sigma_v^2} \right)
    \label{eq:Gaussian}
\end{equation}
The Gaussian $\mathcal{G}(v-V(r),\sigma_{v})$ represents the broadening of the \HI line in velocity space due to internal motions of gas at a given point in the galaxy. The velocity dispersion due to internal motion is taken as $\sigma_v=10$ km/s. We integrate the \HI density at velocity $v$ from every point in the galaxy. In the integral, each point contributes a Gaussian in velocity space centred at its local rotation velocity $V(r)$ weighted by the HI density there.

We corrected the normalisation of the flux density $S(v)$ by integrating it over all channels to obtain the flux. The flux can be scaled with the luminosity distance $D_L$ to get the total \HI mass, which is given as,
\begin{equation}
    \frac{M_{\rm \HI}}{\rm M_{\odot}} = \frac{2.356\times10^{5}}{1+z}\,\bigg(\frac{D_L}{\rm Mpc}\bigg)^2\,\int \frac{S(v)dv}{\rm Jy.km/s}
\end{equation}
Once the normalisation is fixed, the integral in \Eref{eq:HIline} produces a well-representative, symmetric double-horn profile for unlensed mock sources.

For a given \HI mass and source redshift, as we can see from \Eref{eq:signal}, the unlensed signal and SNR also depend on the inclination~($i$) as it affects the linewidth. A face-on source has a narrower linewidth and a sharper peak flux, resulting in a higher SNR compared to an edge-on source. \Fref{fig:Incl_effect} shows the constant SNR contour lines for unlensed face-on~($i=0^\circ$) and edge-on~($i=85^\circ$) sources assuming a $100$-hour uGMRT integration. Since we are assuming a 2D circular profile for our \HI source, having~$i=90^\circ$ for an edge-on source is not feasible. Hence, we assume~$i=85^\circ$ for edge-on sources. For fixed redshift and \HI mass, edge-on sources require relatively higher \HI magnification to reach a given SNR threshold compared to face-on sources. Therefore, face-on sources are easier to detect and can be great targets for \HI detection at higher redshifts with gravitational lensing. We note that \Fref{fig:Incl_effect} is constructed assuming optimal SNR for the underlying source. However, in actual observation, the lack of knowledge about the \HI source properties may result in a loss of some SNR. A more observationally motivated method to calculate the SNR of lensed sources is discussed in the following subsection.

%%%%%%%%%%%%%%%%%%%%%%%%%%%%%%%%%%%%%%%%%%%%%%%%%%%%%%%%%%%%%%%%%%%%%%%%%
\begin{figure*}
    \centering
    \includegraphics[width=0.85\linewidth]{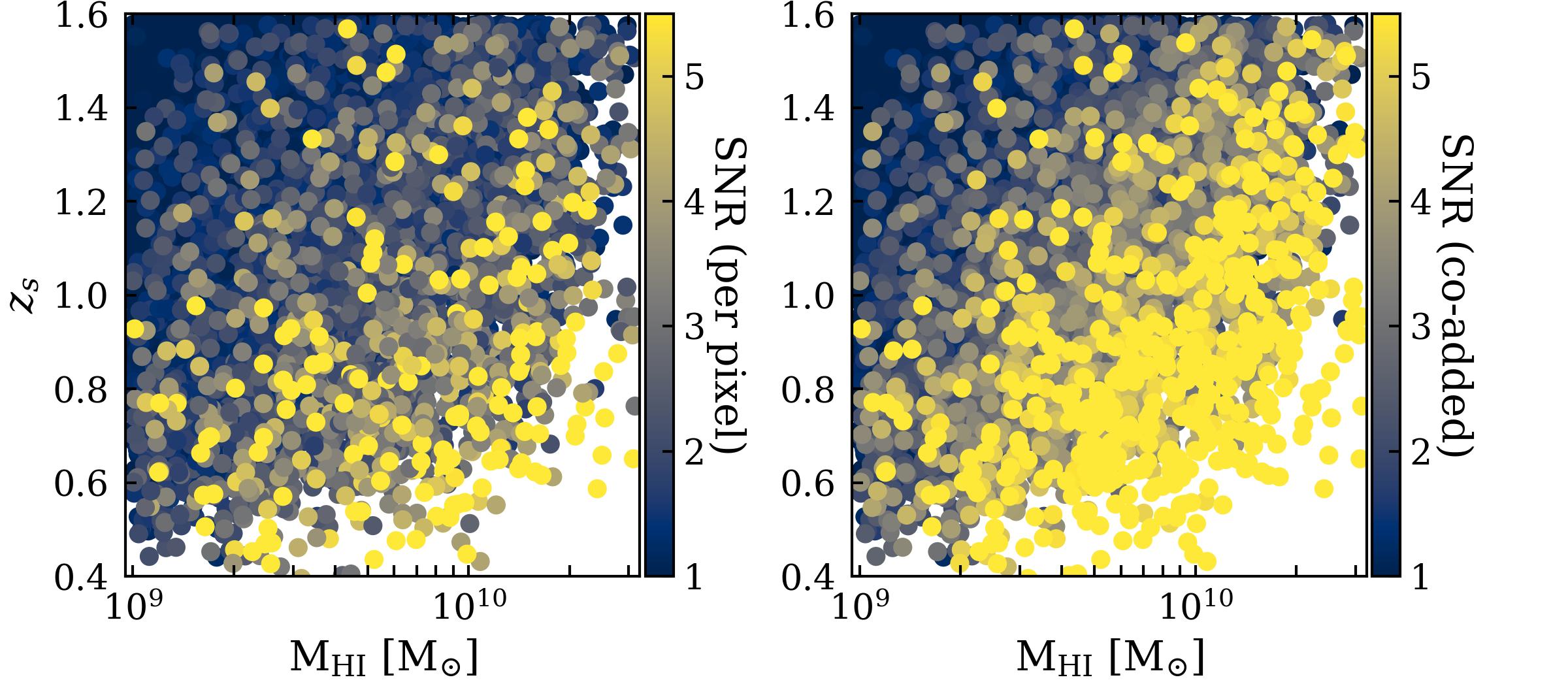}
    \includegraphics[width=0.85\linewidth]{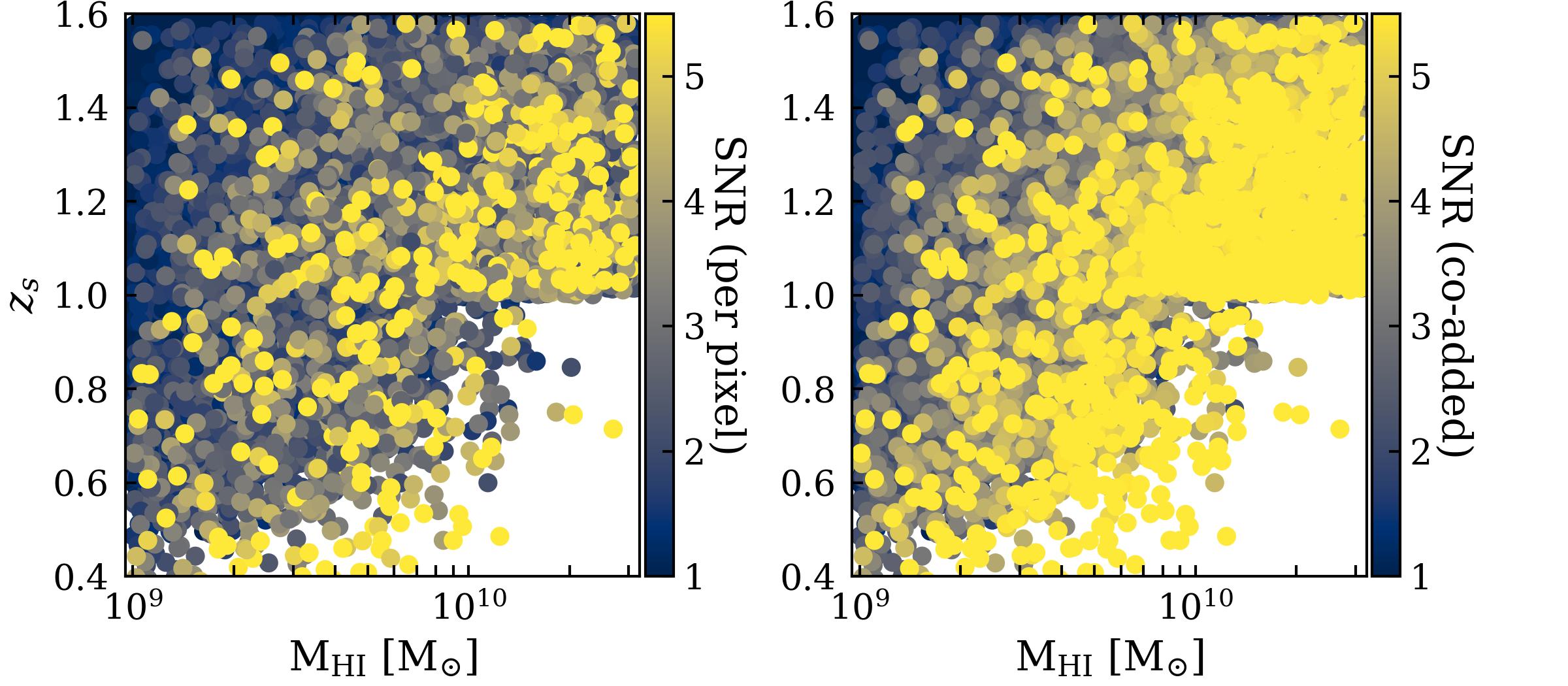}    
    \caption{\HI mass vs. redshift distribution for simulated sources across 200 realisations per cluster (total simulations runs $=200\times50$). The top and bottom rows correspond to ALFALFA and combined \HIMFs, respectively. In the left column, each point represents one lensed source and is colour-coded according to maximum pixel SNR. Each point is colour-coded by its SNR, with higher-SNR points plotted on top to show the expected number and distribution of high-SNR sources for a given redshift and mass. In the right column, the colour represents the total SNR calculated by co-adding all pixels with SNR~$>1$.}
    \label{fig:Scatter}
\end{figure*}

%%%%%%%%%%%%%%%%%%%%%%%%%%%%%%%%%%%%%%%%%%%%%%%%%%%%%%%%%%%%%%%%%%%%%%%%
\subsection{Lensed source SNR}
\label{ssec:lensed_snr}
To compute the SNR for lensed sources, we consider a slightly different approach. We start by assuming that we are searching for lensed sources in a blind survey. Since we do not know the angular extent of the lensed source, we include all available baselines. If the angular size of the source is larger than the uGMRT pixel size, i.e., the angular resolution of the telescope, the source will be resolved into multiple pixels.

In practice, once we determine the position and redshift of mock sources (see \Sref{sec:mock_sim} for more details), we identify all lensed sources behind a given cluster that are either multiply imaged or have magnification~$\geq5$. At this stage, we have assumed background sources to be point sources, and the angular resolution is set equal to the native lensing maps. The above criterion filters out all the sources that will not be sufficiently magnified and will remain below the detection threshold. Next, we determine the uGMRT resolution for the underlying source using \Eref{eq:src_size}. After that, assuming the lens map origin as our centre (i.e, RA, Dec of the cluster lens), we create coarser grids in the image plane with the uGMRT pixel size. Coarse gridding is applied only in the image plane (not in the source plane), as we aim to create SNR maps there. Each pixel on the coarser grid will encompass multiple pixels from the native grid.

As discussed in \Sref{sec:mock_sim}, we have modelled 2d \HI mass density and velocity field at pixel size of native lensing maps (see column 7 of the \Tref{tab:clusters}). However, the \HI intensity and velocity field observed by the telescope are at much poorer resolution, depending on the baselines used and the source redshift. The lensed spectral line (at any coarser resolution) can be reconstructed from the lensed velocity field and \HI density map. In lensing maps, we use a ray-tracing algorithm where each pixel of the velocity field gets mapped from the source to the image plane. A single coarser pixel in the image plane may have contributions from multiple native pixels in the source plane. We co-add all such native pixels within the coarser pixel of the image plane (i.e., uGMRT resolution) with flux weighting. In the image plane, to construct the observed velocity field, each velocity feature is added separately with flux weighting, as we have the spectral line for each native pixel. Therefore, the observed lensed \HI distribution and the \HI velocity field, which a radio telescope measures, are quite different compared to those of an unlensed source. For a velocity bin $(v,v+dv)$, the profile at each coarser pixel is the sum of all native pixels within it and which is given as, 
\begin{equation}
    S(x,y,v) = \sum_i^N\,  M_{HI}(x_i,y_i)\, \mathcal{G}(v-V(x_i,y_i),\sigma_{v}),
    \label{eq:Sv_pixel}
\end{equation}
The flux density $S(x,y,v)$ is converted to the flux (Jy.km/s) by scaling it with the luminosity distance and total H\textsc{i} mass. As the flux is distributed across many velocities, we estimated the integrated flux $F_{\rm HI}(x,y)$ (moment-0 map) by integrating over velocity space and given as,
\begin{equation}
    F_{\rm HI}(x,y) = \int S(x,y, v)\,dv
    \label{eq:FHI}
\end{equation}
The observed velocity field (moment-1 map) at the coarser pixel can be given as follows,
\begin{equation}
    V_{obs}(x,y) = \frac{\int v\,\it S(x,y, v)\,dv}{\int S(x,y, v)\,dv}
    \label{eq:Vobs}
\end{equation}
The integrated flux $F_{\rm HI}(x,y)$ represents the total strength of the H\textsc{i} emission in a pixel. The SNR is calculated as the ratio of the mean flux density to the rms noise per channel, scaled by the square root of the number of independent channels spanning half the width of the signal. Since the outer channels mainly contain weak wings that add noise, only half the linewidth is used for the independent number of channels.
The approximate SNR per pixel can be given as~\citep{Saintonge_2007},
\begin{equation}
    S/N \approx \frac{\beta(\theta)\big(F_{\rm HI}(x,y)/W_{20}\big)}{\sigma_{rms}}\,\bigg(\frac{W_{20}/2}{ d v}\bigg)^{1/2}
    \label{SNR}
\end{equation}
where $W_{20}$ is the linewidth of the unlensed source at $20\%$ peak of the signal, $\beta(\theta)$ is primary beam contribution in the given direction, $dv$ is the channel width ($20$ km/s for uGMRT), and $\sigma_{rms}$ is the rms noise per channel at the uGMRT resolution. The quantity in the square root is the number of channels over which half of the signal is spanned.

%%%%%%%%%%%%%%%%%%%%%%%%%%%%%%%%%%%%%%%%%%%%%%%%%%%%%%%%%%%%%%%%%%%%%%%%
\begin{figure*}
    \centering    
    \includegraphics[width=0.48\linewidth]{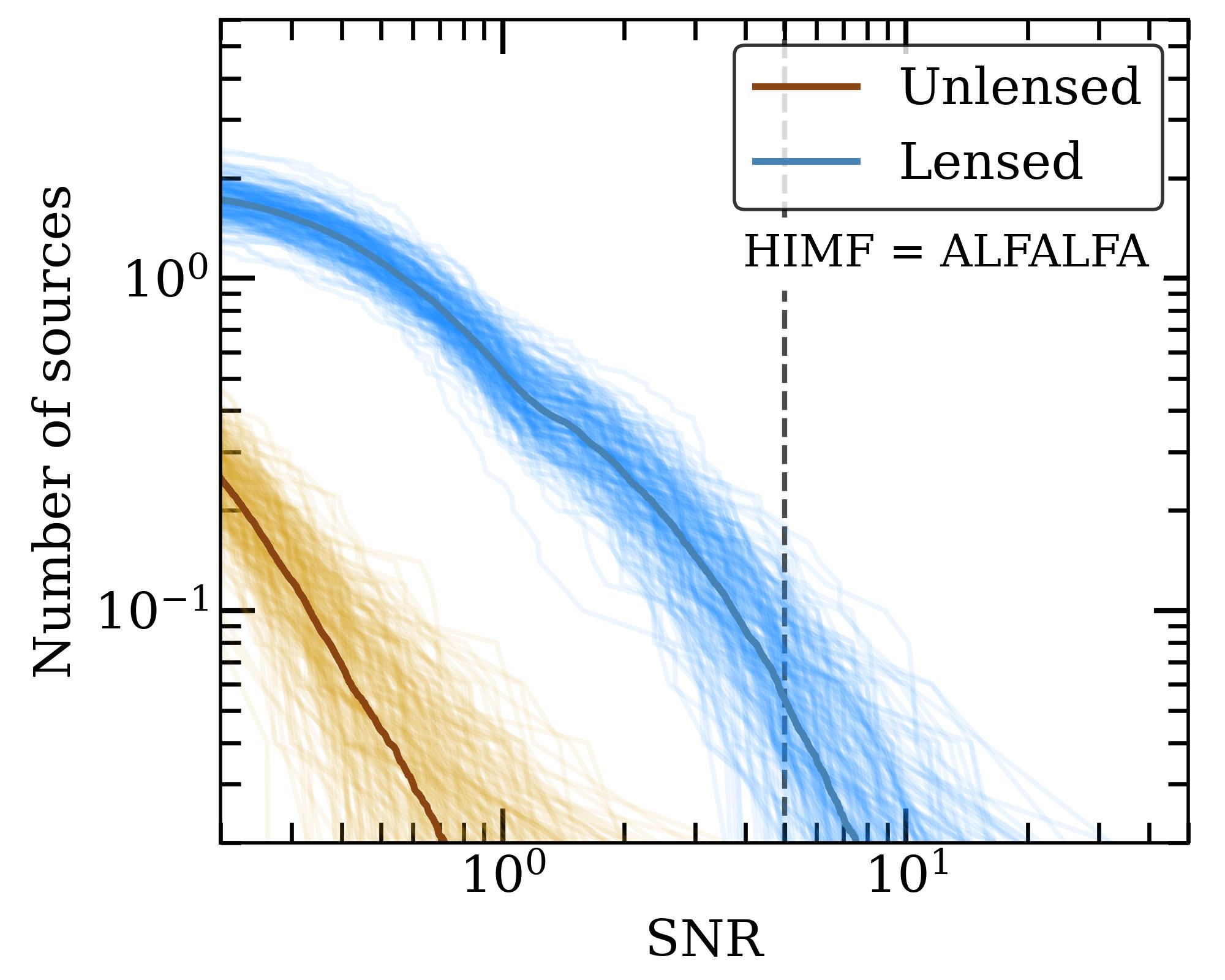}
    \includegraphics[width=0.48\linewidth]{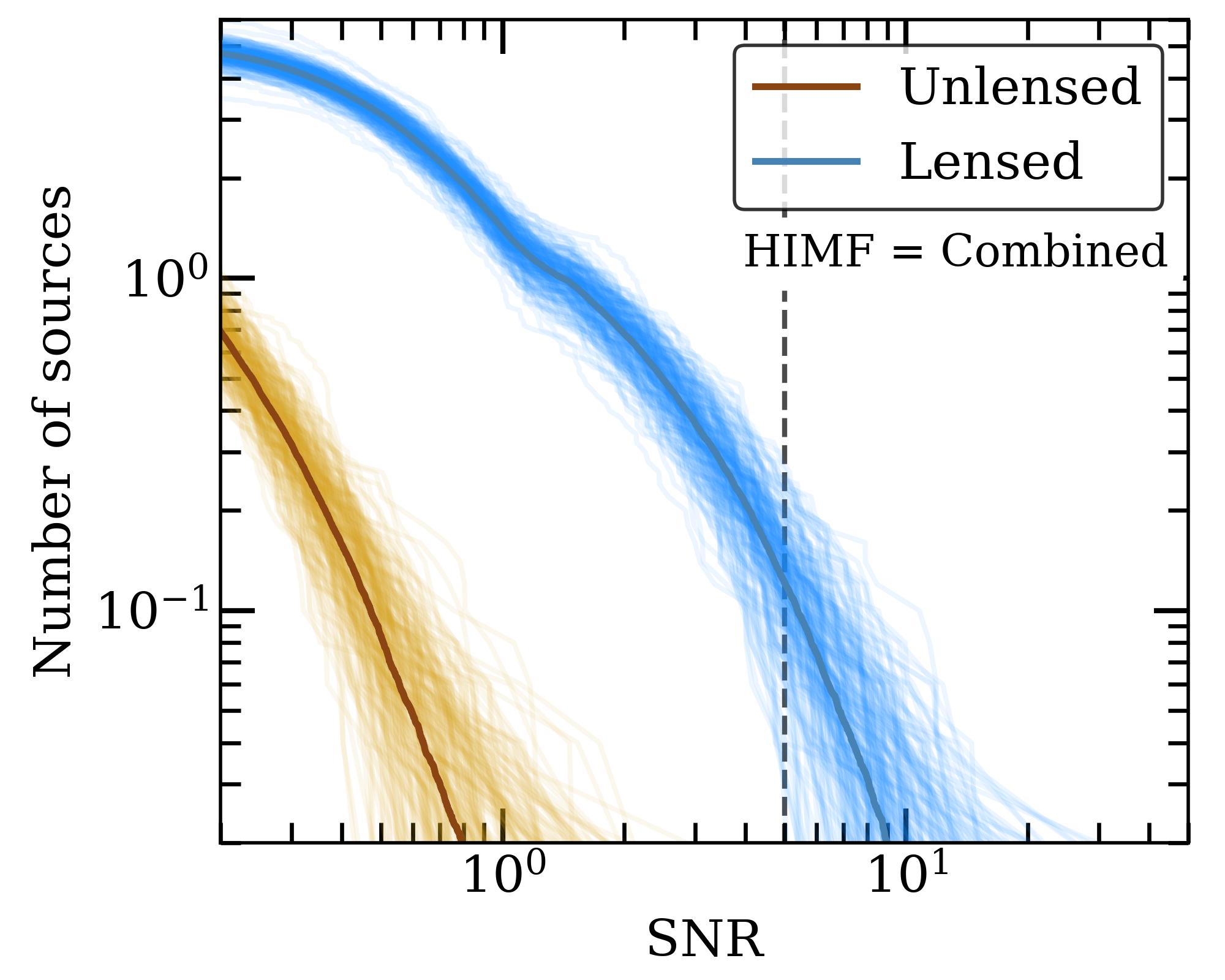}
    \caption{Average cumulative number of (un)lensed sources as a function of SNR. The left and right panels correspond to the ALFALFA and combined \HIMFs, respectively. For each cluster, we simulate 200 realizations, and the thin blue and orange curves represent the number of lensed and unlensed sources per cluster in each realization, respectively. This gives us 200 thin curves. The thick blue and orange curves represent the average of these 200 thin curves for lensed and unlensed cases, respectively. The dashed vertical line mark the $\rm SNR>5.0$~threshold.}
    \label{fig:CDF}
\end{figure*}
%%%%%%%%%%%%%%%%%%%%%%%%%%%%%%%%%%%%%%%%%%%%%%%%%%%%%%%%%%%%%%%%%%%%%%%%

An example of a lensed \HI source in Abell~370 is shown in \Fref{fig:snrmaps}. The top-left panel shows the source plane with caustics and the unlensed source. The central dark region depicts the stellar density profile, and the surrounding orange disk marks the \HI density profile. The \HI velocity profile is shown in the inset plot. The image plane is shown in the top-right panel, and we see that most of the parts of the source are triply imaged, and only a small part is quintuply imaged. In the top row, the angular resolution is equal to the lensing map resolution, which is~$0.05''$ for Abell~370. The lensed 2D \HI velocity distribution is shown in the middle-left panel at the native lensing map resolution, and the middle-right panel shows the \HI velocity profile for each lensed image. Although parts of the source are imaged five times, in the image plane, we only have three isolated images as pairs of images are merging together, explaining the presence of only three lensed velocity profiles in the middle-right panel. To create the velocity profiles shown in the middle-right panel, we assumed a channel width of 8~km/s, which is taken so to capture the smallest velocity features present in the disk which are set by a velocity dispersion of 10~km/s. We can see that since the global minimum image is not distorted (or magnified) significantly, the corresponding line profile~(red curve) is very similar to the unlensed line profile~(black dashed curve). The line profile in blue corresponds to the pair of images that are enclosed by the blue dashed curve in the top-right panel. Here, we observe an additional peak at~$\sim50$~km/s, highlighting the lensing-induced features in the line profile. 

The bottom-left panel in \Fref{fig:snrmaps} shows the sky as seen by the uGMRT. Since the uGMRT resolution is~$3''-5''$ for a source at~$z\sim1$, we only see a handful of pixels belonging to the lensed images, and each pixel is colour-coded according to its SNR. The corresponding observed \HI line profiles are shown in the bottom right panel. A channel width of 20~km/s is assumed, which is a typical velocity resolution and sufficient for detecting broad line emission.  For the red curve, as it corresponds to the global minimum image, we see that again the overall line width is the same as the unlensed source. However, compared to the middle-right panel, we also see an additional peak in the velocity profile at zero velocity due to a high \HI density contribution while summing over native pixels in the central coarser pixel. This central peak is an artifact introduced by the coarser pixelation at the uGMRT resolution. The loss of spatial detail during moment map down-sampling~(at a coarser resolution) introduces artificial jumps in velocity amplitude from one pixel to the other. As we move away from the centre, the \HI density gradually decreases, resulting in a dip around velocities of $\pm30$ km/s. Further from the centre, the \HI density drops more steeply; however, the larger number of pixels at velocities near $\pm70$ km/s contribute to two smaller peaks in the flux density. Note that the line profiles in the bottom-right panel represent what would be derived from the velocity and \HI density maps at the uGMRT spatial and velocity resolution (assuming full bandwidth and maximum possible number of frequency channels), whereas the underlying lensed line profiles are shown in the middle-right panel. Artifact position and amplitude may vary slightly with the pixelization and smoothing strategies used in actual observations.

%%%%%%%%%%%%%%%%%%%%%%%%%%%%%%%%%%%%%%%%%%%%%%%%%%%%%%%%%%%%%%%%%%%%%%%%
\begin{figure*}
    \centering    
    \includegraphics[width=0.8\linewidth]{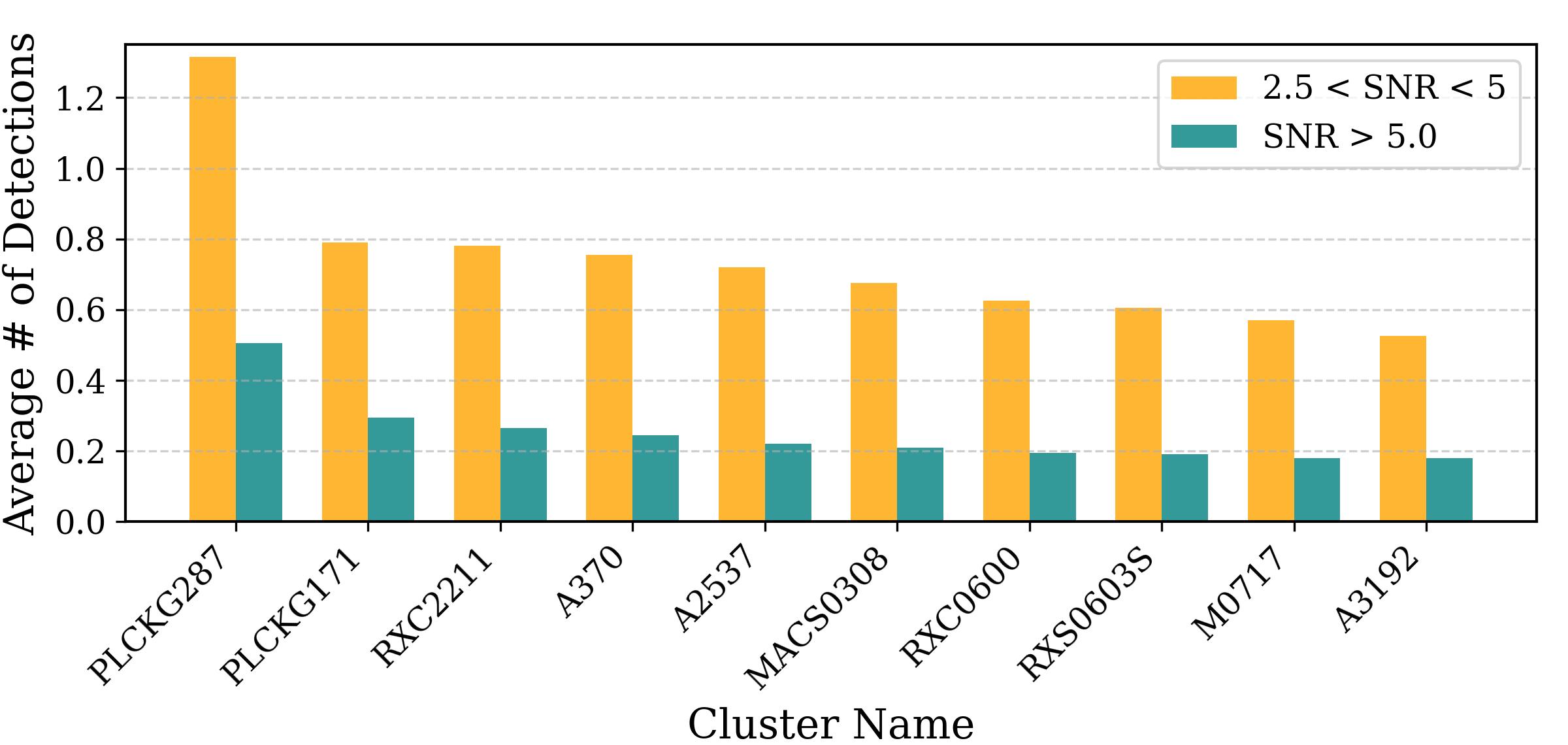} \\
    \vspace{0.5cm}
    \includegraphics[width=0.8\linewidth]{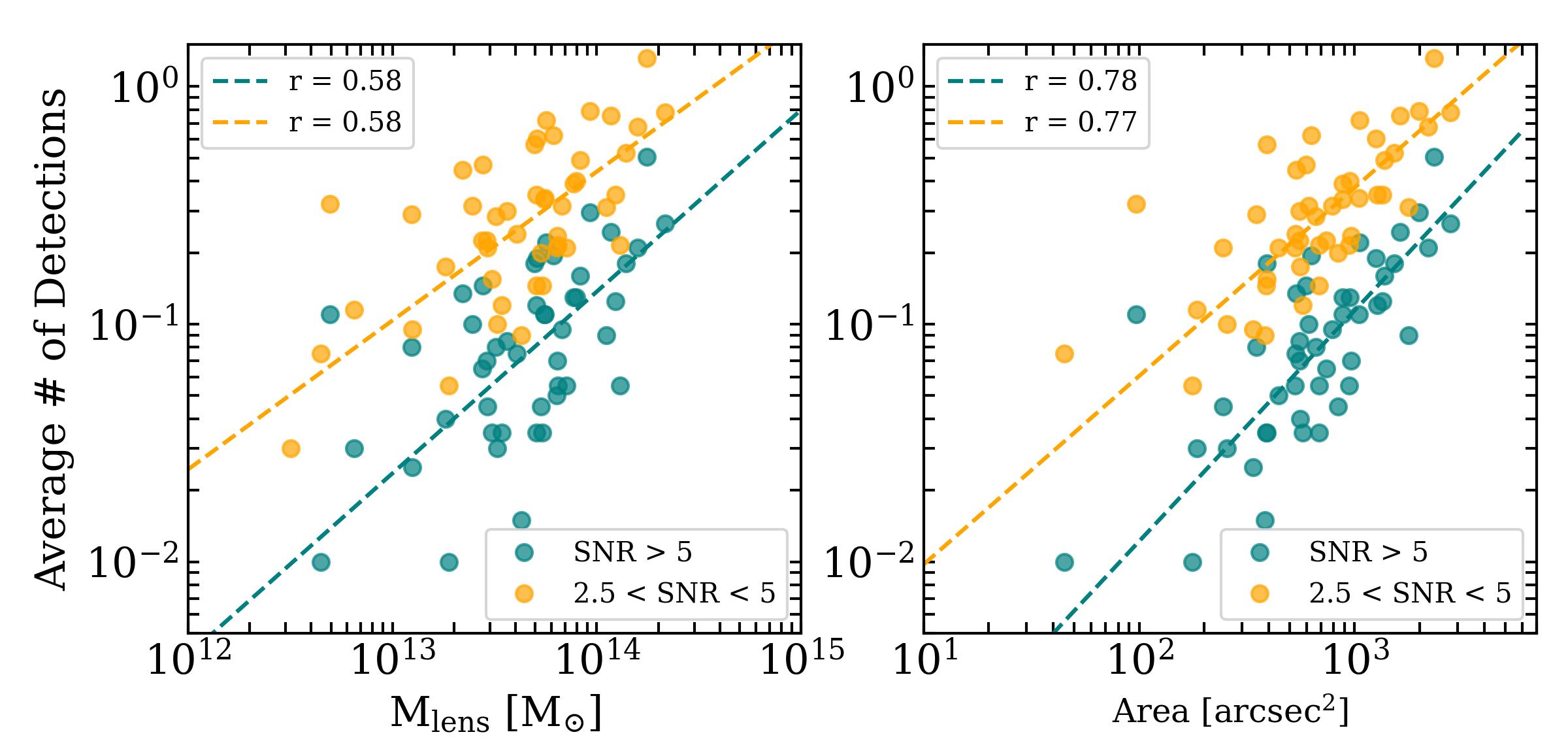}
    \caption{Detection likelihood (i.e., average number of expected detections with at least one source with a certain SNR threshold) for individual clusters. The \emph{top panel} shows the top ten clusters with the highest average number of detections. The green bars correspond to the SNR threshold of 5, whereas the orange bar represents the average number of detection for sources with~$\rm 2.5 < SNR < 5$. Note that the distribution of the number of sources for high SNR matches the Poisson distribution and hence the variance is same as the mean. The bottom-left and bottom-right panels represent the average number of detection as a function of lensing mass and area covered by critical curves at~$z_s=1$ for each cluster in our sample. The same coloured dashed lines are the linear regression fits with the correlation coefficient indicated in the legends.}
    \label{fig:detection_prob}
\end{figure*}
%%%%%%%%%%%%%%%%%%%%%%%%%%%%%%%%%%%%%%%%%%%%%%%%%%%%%%%%%%%%%%%%%%%%%%%%

%%%%%%%%%%%%%%%%%%%%%%%%%%%%%%%%%%%%%%%%%%%%%%%%%%%%%%%%%%%%%%%%%%%%%%%%
\begin{figure*}
    \centering
    \includegraphics[width=0.8\linewidth]{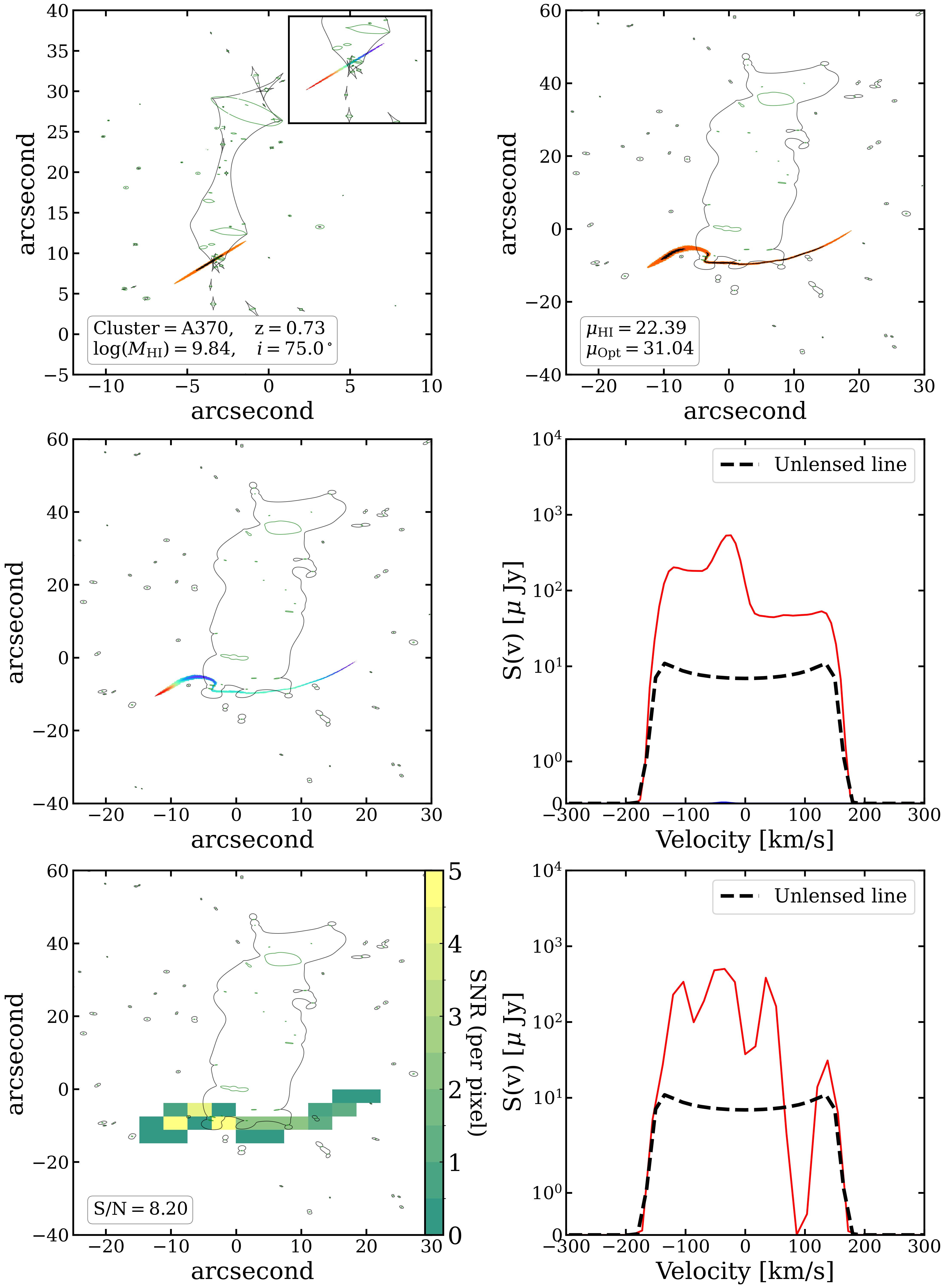}
    \caption{Lensed \HI map in the Dragon Arc in Abell 370 cluster. We refer readers to the caption of \Fref{fig:snrmaps} for a description of various panels.}
    \label{fig:DragonArc}
\end{figure*}
%%%%%%%%%%%%%%%%%%%%%%%%%%%%%%%%%%%%%%%%%%%%%%%%%%%%%%%%%%%%%%%%%%%%%%%%

%%%%%%%%%%%%%%%%%%%%%%%%%%%%%%%%%%%%%%%%%%%%%%%%%%%%%%%%%%%%%%%%%%%%%%%%
\section{Results}
\label{sec:results}

%%%%%%%%%%%%%%%%%%%%%%%%%%%%%%%%%%%%%%%%%%%%%%%%%%%%%%%%%%%%%%%%%%%%%%%%
\subsection{Optical and \HI magnification}
\label{ssec:opt_vs_HI}
In general, \HI sources are expected to have less magnification than their optical counterparts, as larger sources are less magnified. However, as clusters are complex lenses, leading to a large variety of image geometries and may also give rise to image formations where the \HI magnification~($\mu_{\rm \HI}$) is larger than the corresponding optical magnification~($\mu_{\rm opt}$). In \Fref{fig:mu_distribution}, we show the scatter plot of the total \HI vs optical magnification for all of our simulated sources, with each point colour-coded according to their total SNR. The higher-SNR points are plotted on top of the lower-SNR ones to illustrate the expected number and distribution of high-SNR sources. The dashed line represents~$\mu_{\rm \HI} = \mu_{\rm opt}$. We note that sources with $\mu_{\rm \HI} < 4.0$ are predominantly associated with low SNR in the range $[0.5,\,3.5]$, except for a few cases with very high \HI masses at low redshifts $ z = 0.4$. In our simulations of all clusters, the average number of sub-threshold (SNR < 5) sources is $\sim 250$, thus we have around five such sources per cluster.  As we can see in~\Fref{fig:detection_prob}, even the most promising clusters typically have only one such source with SNR in the range $2.5-5$.  Thus, stacking is not a very promising direction for this redshift range.  Observing redshifted $21$~cm line from optical lensed galaxies seems a promising approach towards the first \HI detections in cluster lenses. In the \emph{upper-left} panel, the conditional probability density distribution $P(\mu_x|\mu_y)$ of \HI and optical magnification is shown. The \emph{black solid line} shows the probability density distribution of \HI magnification for sources with $\mu_{\rm opt} > 5$, while the \emph{blue dashed line} represents probability density distribution of the optical magnification for sources with $\mu_{\rm HI} > 5$.

In the scatter plot, we also note that typically, we expect to have lensed sources with similar values for optical and \HI profiles. However, there can be cases where $\mu_{\rm opt}$ is an order of magnitude higher than $\mu_{\rm \HI}$, i.e., points for which~$(\mu_{\rm \HI}, \mu_{\rm opt})\sim(10, 10^2)$, although for such points the SNR is~$\lesssim3$. Although less, we also have cases where~$\mu_{\rm \HI} > \mu_{\rm opt}$. To gain insight into such cases, we show source plane cutouts of four such cases in the right part of \Fref{fig:mu_distribution}. We note that in all of the cases, the source orientation is parallel to the caustics, with a large part of the \HI profile lying in the high magnification region. Looking at cutouts ~(a) and~(d), we can see that it is not required to have a complex caustic network such that~$\mu_{\rm \HI} > \mu_{\rm opt}$ and even a relatively simple caustic structure, such as for an elliptical lens, is sufficient. In Fig.~\ref{fig:high_muHI} of the appendix \ref{sec:appendix} we present a more detailed view of three representative cutouts, along with their corresponding lensing features in the image plane.

%%%%%%%%%%%%%%%%%%%%%%%%%%%%%%%%%%%%%%%%%%%%%%%%%%%%%%%%%%%%%%%%%%%%%%%%
\subsection{\HI detection in cluster lenses}
\label{ssec:HIdetection}
We simulate $100$-hour mock observation with uGMRT to detect \HI emission for each cluster lens. For each of the 50 cluster lenses, we generate 200 realisations to obtain the lensed source population for each of ALFALFA and combined mass functions. Here we present the expected outcome for $50$ cluster lenses (see \Tref{tab:clusters}), the results from a total of $50 \times 200$ simulations. The corresponding mass vs. redshift plots are shown in \Fref{fig:Scatter}. The top and bottom rows correspond to ALFALFA and combined \HIMF~(described in \Sref{sec:mock_sim}), respectively. In the left column, we color-coded the pixels according to the maximum SNR among all lensed pixels for each source, whereas in the right column, we show the total co-added SNR from all lensed pixels with SNR $>1$ for each source. In all panels, higher-SNR points are plotted on top of lower-SNR ones to illustrate the expected number and distribution of high-SNR sources in the redshift-mass plane. For the ALFAFLA case (top-left panel), we see a relatively low number of yellow points, implying that only a small number of sources are capable of leading to individual pixels with SNR $>5$. With co-addition (top-right panel), the SNR increases with the square root of the number of pixels, i.e. $\sqrt{N_{\rm pix}}$, if each pixel has the same per-pixel SNR. Therefore, we see an increase in the number of detected sources. It is important to note that, even without co-adding, we have sources with pixel SNR~$>5$, although few, all the way up to~$z_s\simeq1.6$, and co-adding the SNR only increases the number of such sources. Compared to ALFALFA, the combined \HIMF leads to more sources with SNR~$>5$, especially at~$z_s\geq1.0$ as the underlying mass function~\citep{Chowdhury_2024} leads to a higher number density of massive sources in the same redshift range. \Fref{fig:Scatter} shows that in a blind mock \HI survey of galaxy clusters, some high-significance lensed \HI detections can be made all the way up to~$z_s\simeq1.6$.

It is important to note that the combined \HIMF is expected to be more representative of the \HI mass distribution of the source population compared to the ALFALFA \HIMF, as it accounts for the measurements around redshift $z \approx 1$. That said, these measurements are based on \HI stacking, which implies that our predicted number of detections is conservative. As we have noted in the Fig.~\ref{fig:Incl_effect}, the contour lines for the face-on sources with SNR $3$, $5$, and $7$ are extended diagonally from bottom to top. In the scatter plot, we also see a similar behaviour where sources with high SNR (yellow points) lie in a diagonal region. These sources are mostly face-on with sharp peak fluxes. Although the inclination effect on SNR in Fig.~\ref{fig:Incl_effect} is shown without considering lensing, this effect also remains significant in the presence of lensing of \HI emission.

In \Fref{fig:CDF}, we plot the cumulative distribution of the average number of sources per cluster in the sample of 50 clusters as a function of co-added SNR. The left and right panels correspond to ALFALFA and the combined \HIMFs, respectively. The orange and blue curves correspond to unlensed and lensed number counts. Since we have 200~realisations for each cluster, we have 200~thin lines for both lensed and unlensed sources. The thick solid line is the average of the thin lines. The black vertical dashed line indicates the $\rm SNR = 5$ threshold. Again, as expected, we see larger source counts for the combined mass function than ALFALFA due to a higher number density of sources at~$z_s\approx1$. Lensing magnification also provides a more than two orders of magnitude to the source number counts at~$\rm SNR=1$ and even more at~$\rm SNR=5$. From \Fref{fig:CDF}, we expect to detect $1-5$ lensed \HI sources in the sample of 50~clusters, assuming the ALFALFA mass function, where the combined \HIMF gives~$3-13$ lensed sources. With $100$~hrs of uGMRT observation, the average number per cluster is one with $\rm SNR \simeq 1.7$. To push this to the detection threshold of $\rm SNR \gtrsim 5.0$, roughly 900~hrs of uGMRT observation (per cluster) is required, as the SNR varies as $\sqrt{\Delta t}$ in the ideal case.

%%%%%%%%%%%%%%%%%%%%%%%%%%%%%%%%%%%%%%%%%%%%%%%%%%%%%%%%%%%%%%%%%%%%%%%%
\begin{figure*}
    \centering
    \includegraphics[width=0.8\linewidth]{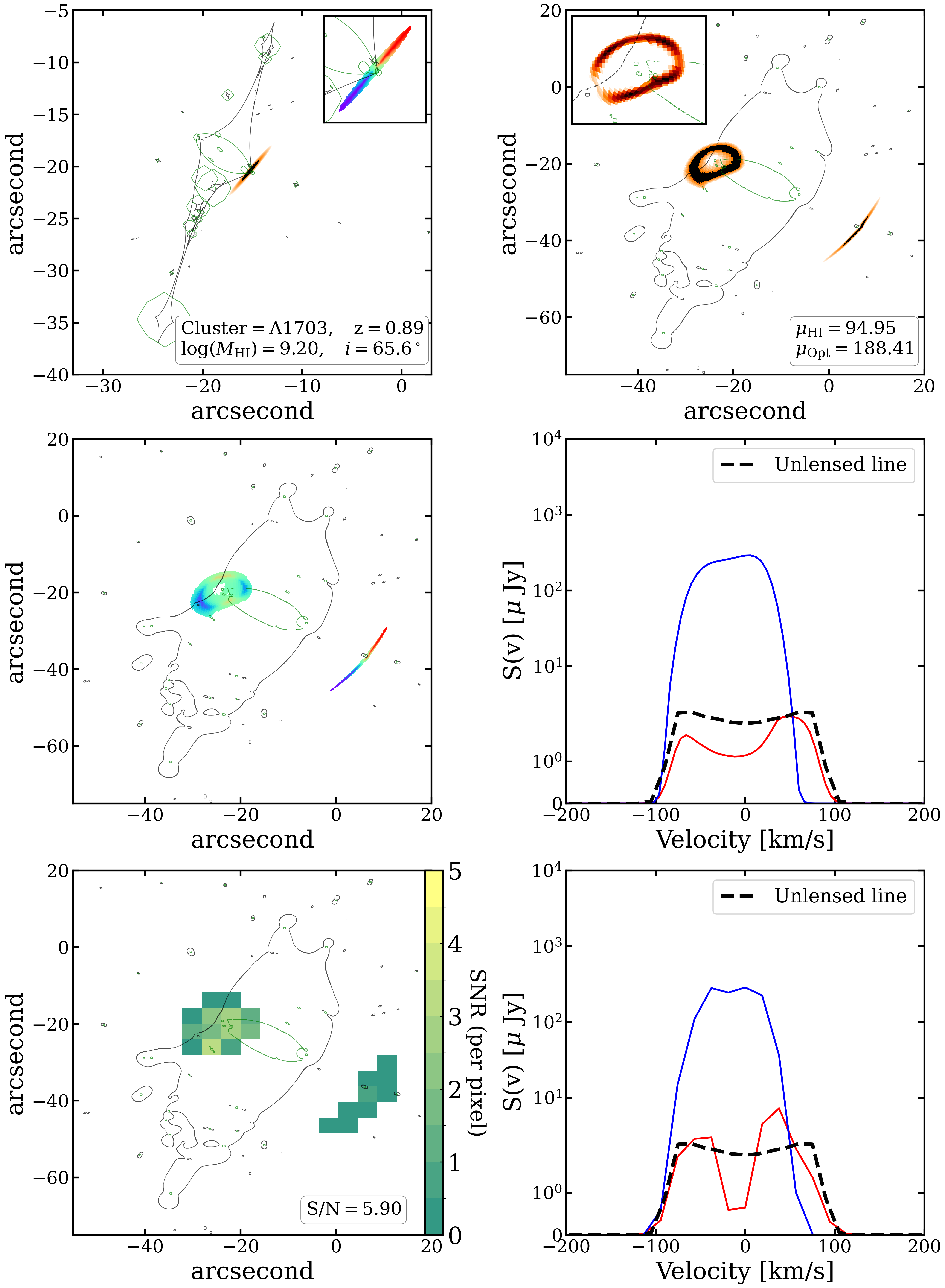}
    \caption{Lensed \HI map in the HU image formation in Abell 1703 cluster. We refer readers to the caption of \Fref{fig:snrmaps} for a description of various panels. The inset plot in the top right panel shows the zoomed-in view of HU image formation.}
    \label{fig:HU}
\end{figure*}
%%%%%%%%%%%%%%%%%%%%%%%%%%%%%%%%%%%%%%%%%%%%%%%%%%%%%%%%%%%%%%%%%%%%%%%%

%%%%%%%%%%%%%%%%%%%%%%%%%%%%%%%%%%%%%%%%%%%%%%%%%%%%%%%%%%%%%%%%%%%%%%%%
\subsection{Detection likelihood in individual clusters}
\label{ssec:individul}
In this section, we primarily ask: Do some lensing clusters give a higher likelihood for \HI detection? The likelihood that a given cluster lens is more efficient in leading to a higher number of lensing \HI sources is directly proportional to the average number of expected detections that lead to at least one lensed \HI source with SNR above a given threshold. Hence, we relate the \textit{detection likelihood}~to the average number of expected detections in which the SNR (for at least one source) exceeds a given threshold. If the average number of expected detections exceeds unity, it implies that, for a given detection threshold, more than one source is detected per realisation. In the top panel of \Fref{fig:detection_prob}, we show the detection likelihood for the ten most efficient lensing clusters in our sample. The orange and green bars illustrate detection likelihood with $2.5 < \rm SNR< 5.0$ and $\rm SNR> 5.0$, respectively.

We note that most of the clusters have masses~$\gtrsim5\times10^{14}\:{\rm M_\odot}$~\citep[e.g.,][]{2020ApJ...889..189S} leading to a relatively larger lensing cross-section. The effect of lensing mass (enclosed within the critical curves at~$z_s=1$) on the average number of detection is further shown in the bottom-left panel of \Fref{fig:detection_prob}. We observe that an increase in the lensing mass leads to an increase in the average number of detection, which is further highlighted by the positive correlation coefficient~($r\simeq0.6$) mentioned in the plot legend. The bottom-right panel shows the average number of detection as a function of the area enclosed by critical curves at~$z_s=1$. Here we have an even stronger positive correlation~($r\simeq0.8$). This seems to indicate that the presence of substructures is correlated with a higher chance of detection, as these lead to a larger area in the source plane with a high lensing magnification.

Looking at the green bars for the top-three galaxy clusters, the corresponding average number of detections imply that we would detect one lensed \HI source in one of these clusters. Hence, it would be interesting to check whether these lensing clusters with high chance of detection actually contain any optical sources with high magnification, as those can be good targets for future follow-up programs for lensed \HI detection. Unfortunately, visual inspection of the top-three clusters did not show any strong candidates at~$z_s\leq1.58$ for future \HI follow-ups. That said, the fourth cluster, Abell~370, contains the well-known ``Dragon arc'', an excellent candidate for lensed \HI detection with current facilities~\citep{Blecher2024}, which we further discuss in the following subsection.

%%%%%%%%%%%%%%%%%%%%%%%%%%%%%%%%%%%%%%%%%%%%%%%%%%%%%%%%%%%%%%%%%%%%%%%%
\subsection{Abell 370 -- Dragon Arc}
\label{ssec:A370}
One of the excellent lensed optical candidates for \HI follow-up observations is Dragon Arc in the Abell~370 galaxy cluster. The source redshift
for this system is ~$z_s=0.7251$. Recent work~\citep{Blecher2024} has shown that MeerKAT can detect \HI signal in this with 50~hours of integration time. In this section, we study the prospects of detecting the \HI signal in this system with uGMRT. We use the source parameters given in \citet{Blecher2024} to simulate an unlensed source with the method described above. The source parameters are, $\log(M_{\rm \HI}/{\rm M_{\odot}}) = 9.84 \pm 0.27$, $i=75.5\pm5^\circ$ and~$\phi = 130^\circ\pm 5^\circ$. For our analysis, we use the \textsc{Lenstool} model as mentioned in \Tref{tab:clusters}. Since only a part of the whole source is multiply imaged, we find the source position corresponding to the optical light centre to determine the position of our simulated source in the source plane.

The results, assuming best-fit parameters for the source, for 100-hour integration time with uGMRT are shown in \Fref{fig:DragonArc}. In our analysis, we get~$\mu_{\rm opt} \sim 31$ and~$\mu_{\rm HI}\sim22$, which are in agreement with previous studies~\citep{2010MNRAS.402L..44R, Blecher2024}. Since only a part of the source is strongly lensed, we see its effect on the lensed line profile, where only the blue-shifted part shows bumps in addition to an overall magnification, as can be seen in the middle-right panel. As mentioned in the bottom-left panel, with 100-hour integration time, we get co-added SNR~$>8$, highlighting an extremely high chance of lens \HI detection in this source in 100 hours. A $5\sigma$ detection could be achieved even with just 40 hours of integration time with uGMRT. Our estimated observing time is consistent with what was predicted by \citet{Blecher2024} for detection with MeerKAT, given that uGMRT and MeerKAT are comparable in terms of sensitivity. The bottom-right panel shows the \HI line profile at the uGMRT spatial and velocity resolutions of $\thicksim 5''$ and $\sim 20$~km/s, respectively. The flux contribution at red-shifted velocities is lower than at blue-shifted velocities. The missing flux at $\approx 100$~km/s due to the pixelization effect (none of the pixels have significant flux corresponding to $\approx +100$~km/s velocity in this case and hence we see a dip in the spectral line) as each velocity feature is added with flux density as a weight to make the observed velocity map at uGMRT resolution (see Eq. \ref{eq:Sv_pixel} and Eq. \ref{eq:Vobs}). 

After the submission of the original version of this manuscript, a detection of the \HI in the dragon arc has been claimed with MeerKAT observations~\citep{2025arXiv251101715L}. This supports our optimistic prediction for this system.

%%%%%%%%%%%%%%%%%%%%%%%%%%%%%%%%%%%%%%%%%%%%%%%%%%%%%%%%%%%%%%%%%%%%%%%%
\subsection{Abell 1703 -- Hyperbolic Umbilic}
The Hyperbolic Umbilic (HU) is a specific image configuration \citep{2020MNRAS.492.3294M, Meena_2023} of four highly magnified images arranged in a ring-like structure off-centred from the cluster centre. One such image formation is observed in Abell~1703 for a source at~$z_s=0.8889$~\citep[e.g.,][]{2008A&A...489...23L}. Due to the four highly magnified images in the systems, such image formations are another great target to detect the lensed \HI signal. In this section, we study the feasibility of lensed \HI detection in the above HU system in Abell~1703. To determine the source parameters, we start by measuring lensing-corrected photometry of one of the lensed images (which is part of the HU image configuration) in HST-ACS/WFC~(F435W, F475W, F555W, F625W, F775W, and F850LP) and HST-WFC/IR~(F125W and F160W). After that, we use \textsc{Bayesian Analysis of Galaxies for Physical Inference and Parameter EStimation}~\citep[\textsc{bagpipes;}][]{2018MNRAS.480.4379C} to infer the stellar mass of the source galaxy. To estimate the orientation of the source on the sky, we estimate the lensing corrected ellipticity and position angle. With the above, we have $\log(M_\star / {\rm M_\odot})=9.0$, $i = 65.6^\circ$, and $\phi= 48.44^\circ$ and use these values to simulate optical and \HI profiles for our source. Using inverse ray shooting from each lensed image position, we determine the corresponding barycentre (i.e., flux-weighted source position) and use it as our source position. 

The results for \HI lensing in Abell~1703 HU image formation are shown in \Fref{fig:HU}. As expected of HU image formations, the lensed images are highly magnified with~$(\mu_{\rm opt}, \mu_{\rm \HI}) = (191, 95)$. Unlike the observed HU image formation in optical, in our simulated case, all four images are merging together, forming a complete ring, which is likely to be due to the larger source size (see \Sref{sec:mock_sim} for more details) or an error in the source position relative to the caustics as the underlying lens model is optimised using a point source.  However, we do not expect this to considerably affect our estimations. The co-added SNR over the entire image plane is estimated to be $5.9$ in $100$ hours of uGMRT integration, and a detection with $5\sigma$ confidence can be achieved in $\approx75$ hours of uGMRT integration. As the four images are merging together, we only have two lensed line profiles in the middle/bottom-right panel of \Fref{fig:HU}. The blue line profile, which corresponds to the ring image formation, shows a significant bump around $-10$~km/s, as in our simulation, the central part of the lens sits in the five-image region in the source plane.

%%%%%%%%%%%%%%%%%%%%%%%%%%%%%%%%%%%%%%%%%%%%%%%%%%%%%%%%%%%%%%%%%%%%%%%%
\section{Conclusions}
\label{sec:conclusions}

Neutral hydrogen, a primary component of galaxies, represents the raw fuel for future star formation and plays a significant role in galaxy evolution. Lack of direct \HI emission detections at intermediate to high redshifts $(z \gtrsim 0.4)$ limits our understanding of its distribution in galaxies. One way to push the redshift limit further with current and future facilities is by observing strongly lensed systems with high magnification factors. In our current work, with a sample of fifty galaxy clusters, we study the feasibility of detecting strongly lensed \HI at~$0.4 < z_s < 1.58$ with uGMRT in 100~hours of integration time. Our main findings are as follows:
\begin{itemize}
    \item We compare the optical and \HI magnifications of the simulated source population. For a typical (i.e., when a source does not lie on a caustic) lensing scenario, optical and \HI magnifications are expected to be similar. However, once we situate the source close to the caustics, we can observe large deviations between \HI and optical magnifications, including cases where \HI magnification exceeds the corresponding optical magnifications.
    
    \item We performed $200$ realisations per cluster and created SNR maps at uGMRT observing resolution, where signal-to-noise ratio is presented per pixel for $100$ hours of integration. The SNR maps~(per-pixel) are notably useful for optimising \HI detection in blind radio observations. We do co-addition to get the full SNR of the source and found that the \HI detections up to $z\simeq1.6$ are possible for galaxies with \HI mass $\gtrsim10^{9} \rm\,M_{\odot}$. This suggests a possible chance of detecting $ 21$~cm emission with strong lensing at such high redshifts.

    \item We study the likelihood of \HI detection for a given detection threshold and found the average number of detection of a $5\sigma$-detection per cluster is very low, $P({5\sigma})\approx 0.11$ (  i.e., on average~$5-6$ sources in 50 clusters are detectable), within $100$ hours of uGMRT integration time. By increasing the integration time to 900~hours per cluster lens, we can achieve, on average, one lensed \HI detection with~$5\sigma$ confidence per cluster. That said, the detection likelihood can also vary significantly across individual clusters, with some clusters having $P({5\sigma})\approx 0.4$.

    \item The average number of \HI detection is positively correlated with the lensing mass and area enclosed by critical curves. A few clusters have more area under critical curves given a redshift, and the likelihood of \HI detection in such clusters is relatively high. For these favourable clusters, our results show that there is a $100\%$ chance of detecting \HI emission with $\approx400$ hours of uGMRT integration.

    \item Focussing on individual systems, we study the prospects of lensed \HI detection in the well-known Dragon Arc in the Abell~370. We find that $40$--$50$ hours with uGMRT are sufficient to achieve a $5\sigma$ lensed \HI detection in this system. Another highly magnified system is the HU image formation in Abell~1703, where a $5\sigma$ detection of lensed \HI signal with uGMRT would require~$\sim75$~hours.

    \item We find that for sources lying close to or across a caustic, lensing can introduce significant distortions to the observed line profile. These distortions can introduce non-uniformities in observed \HI gas density and velocity maps, resulting in very non-trivial spectral lines. The diversity of \HI line profiles arising from strong lensing implies that a simple-minded analysis based on a generic line shape can lead to a suppression of SNR. For systems where we can model the source and the lens, we can make predictions and use matched filtering for effective signal extraction and optimal detection.     
\end{itemize}

Our simulations can be further refined by making the source model more realistic. For example, recent observations~\citep{2025ApJ...984...15Y} have given us invaluable insight into the typical distribution of \HI perpendicular to the plane of disk galaxies. Incorporating such a three-dimensional model for \HI sources can further improve our model and, in turn, our predictions. Considering three-dimensional source models also leads us to more realistic magnification estimates for highly inclined sources compared to two-dimensional sources. In addition, we have used a jump in the \HI mass function to mimic a sharp transition observed using stacking. SKA's pathfinder surveys, such as DINGO~\citep{Duffy_2012} and LADUMA~\citep{2024AAS...24334607B}, will measure the \HIMF to $z < 0.6$, while the SKA will extend this to $z \sim 1$. A smoother model of \HIMF with redshift can be obtained by combining the mass functions over $0.2 \lesssim z \lesssim 1.0$ from these surveys. At higher redshifts ($z \gtrsim 1$), \HIMF\ estimates from enhanced \HI stacking measurements will provide further refinement of our model~\citep[e.g.,][]{Chowdhury2020,Chowdhury2021,Chowdhury_2024}. Lastly, we plan to do \HI-lensing simulations for other telescopes capable of observing the redshifted $21$~cm line at intermediate and high redshifts, and to investigate the lensed optical systems at corresponding redshifts.

%%%%%%%%%%%%%%%%%%%%%%%%%%%%%%%%%%%%%%%%%%%%%%%%%%%%%%%%%%%%%%%%%%%%%%%%
\section*{Acknowledgements}
%Authors thank Nissim Kanekar and Prasun Dutta for useful comments. 

The authors thank Prasun Dutta for his useful comments. The authors also thank the referee, Tariq Blecher, for the useful comments. The authors acknowledge the use of the PARAM Smriti at NABI Mohali and the HPC facility at IISER Mohali for providing computational resources. This research has made use of NASA's Astrophysics Data System Bibliographic Services. SB thanks the Department of Science and Technology (DST), Government of India, for financial support through the Council of Scientific and Industrial Research-UGC research fellowship. The authors also acknowledge Swati Gavas for helpful discussion regarding the mass function. JSB was on sabbatical at NCRA-TIFR during the time when this work was initiated. This work utilizes gravitational lensing models produced by PIs Bradač, Natarajan \& Kneib (CATS), Merten \& Zitrin, Sharon, Williams, Keeton, Bernstein and Diego, and the GLAFIC group. This lens modeling was partially funded by the HST Frontier Fields program conducted by STScI. STScI is operated by the Association of Universities for Research in Astronomy, Inc. under NASA contract NAS 5-26555. The lens models were obtained from the Mikulski Archive for Space Telescopes (MAST). This work is based on observations taken by the RELICS Treasury Program (GO 14096) with the NASA/ESA HST, which is operated by the Association of Universities for Research in Astronomy, Inc., under NASA contract NAS5-26555.

%%%%%%%%%%%%%%%%%%%%%%%%%%%%%%%%%%%%%%%%%%%%%%%%%%%%%%%%%%%%%%%%%%%%%%%%
\section*{Data Availability}
All the data products used in this article are publicly available. Additional data products created in the current work can be easily generated using the methods discussed in the text.

%%%%%%%%%%%%%%%%%%%%%%%%%%%%%%%%%%%%%%%%%%%%%%%%%%%%%%%%%%%%%%%%%%%%%%%%
\bibliographystyle{mnras}
\bibliography{Reference}  

\appendix
\section{High \HI Magnification cases}
\label{sec:appendix}
Here we show the source and image planes for a few cases, presented as cutouts in Fig.~\ref{fig:mu_distribution}, where the \HI magnification is significantly higher than optical magnification.
\begin{figure*}
    \centering    
    \includegraphics[width=0.7\linewidth]{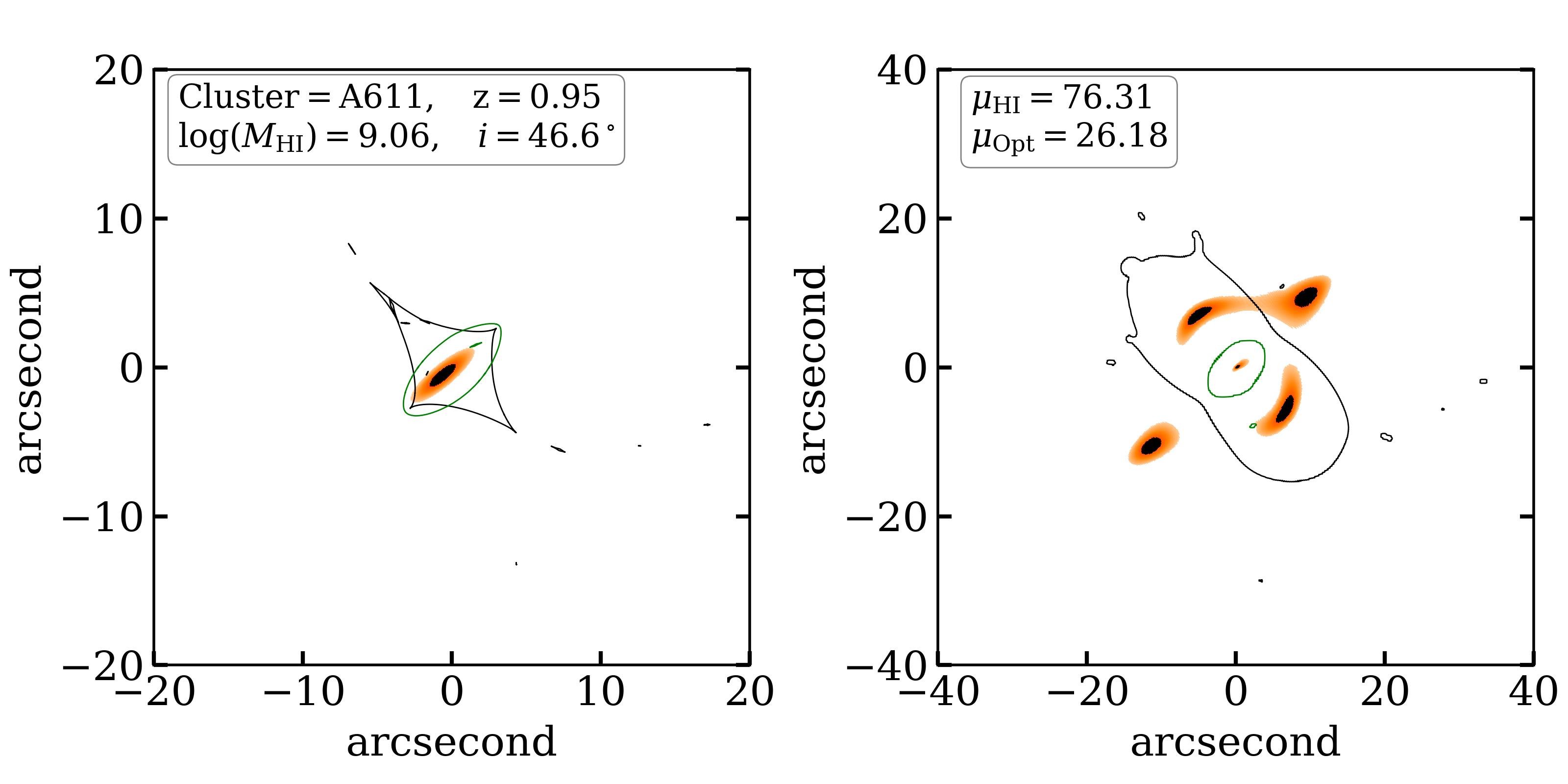} \\
    \vspace{0.1cm}
    \includegraphics[width=0.7\linewidth]{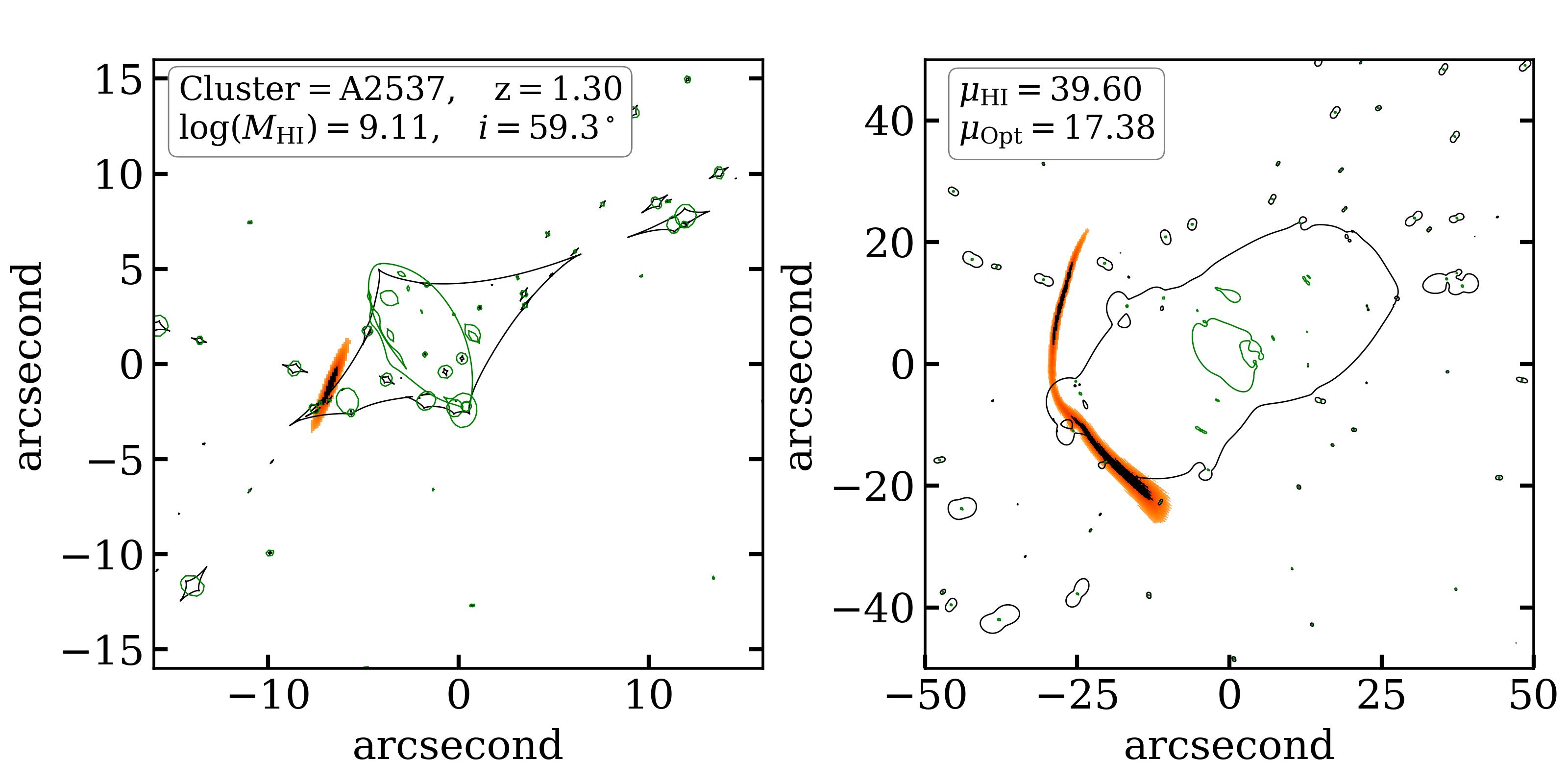}
    \includegraphics[width=0.7\linewidth]{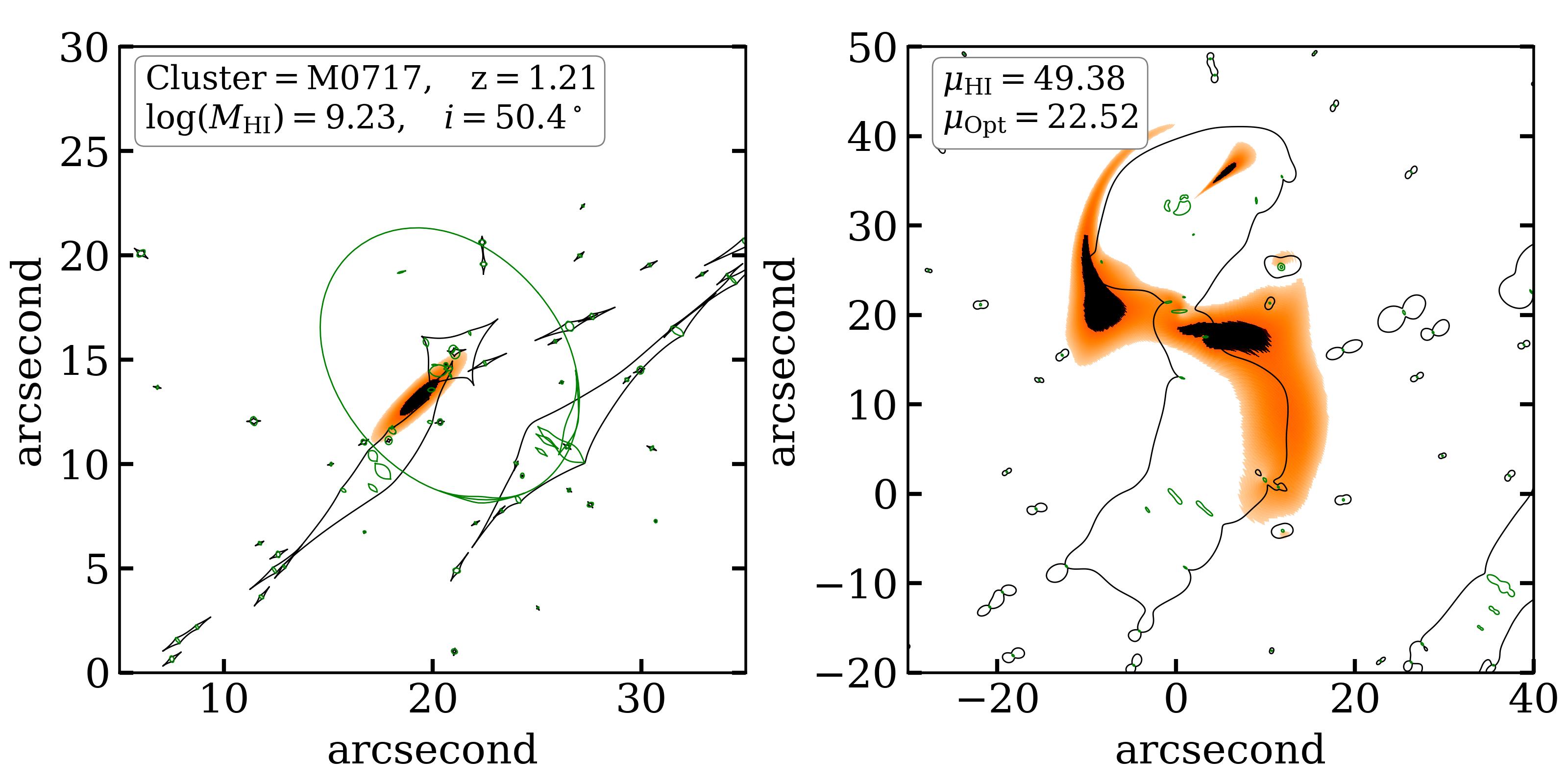}
    \caption{Source and image planes for high \HI magnification cases. The left column shows the source plane, and the right column shows the image plane.}
    \label{fig:high_muHI}
\end{figure*}

%%%%%%%%%%%%%%%%%%%%%%%%%%%%%%%%%%%%%%%%%%%%%%%%%%%%%%%%%%%%%%%%%%%%%%%%
% Don't change these lines
\bsp	% typesetting comment
\label{lastpage}
\end{document}